# Trajectory Functionals in the Thermodynamics of Radiation-Induced Damage in Structural Materials

V. V. Ryazanov, https://orcid.org/0000-0002-5308-3212

Institute for Nuclear Research, pr. Nauki, 47 Kiev, Ukraine, e-mail: vryazan19@gmail.com

This paper delivers a novel self-consistent non-local kinetic framework based on dose-dependent semi-Markov renewal transport and the generalized Doob's h-transform to evaluate the structural degradation of WWER-1000/1200 fuel claddings (Zr–1%Nb alloy). By leveraging the Feynman–Kac formalism, we map the continuous microscopic defect path trajectories generated by the Liouville operator directly onto the evolution of the macroscopically integrated damage potential, Φ. The framework accounts for long-range memory effects in the degradation kinetics via a non-vanishing structural commutator of the path Liouville operator and Zwanzig–Mori projection operators, which mathematically links the spatial restrictions of microscopic sinks to energetic path penalties. Spatial profiling of the cladding's mean lifetime, $\langle \tau \rangle$, demonstrates a distinct sigmoidal, S-shaped degradation trajectory encompassing distinct incubation, linear growth, and avalanche breakdown stages. The numerical results reveal an abrupt, localized reduction in the operational lifespan near the cladding boundaries. This marginal vulnerability is shown to be a direct physical consequence of boundary stress field singularities at the fuel-cladding interface and localized embrittlement. The developed path-integral architecture provides a strict thermodynamic link between non-equilibrium fluctuations and the global Griffiths macro-barrier, offering a robust predictive tool for high-burnup nuclear fuel safety assessments.

## 1. Introduction

Radiation materials science investigates both the thermodynamics of defect creation and their subsequent impact on the macroproperties of metals [1]. The foundations of displacement cascade theory were established in [2]. A detailed analysis of the behavior of defects, voids, and gases in fissile materials under irradiation is provided in [3]. Point defects and their electronic origin are described in [4]. Radiation damage spans vast ranges of scale: from picoseconds and angstroms (the primary knocked-on atom, displacement cascade) to decades and meters (radiation swelling, reactor pressure vessel creep). Consequently, hierarchical multiscale modeling is widely utilized in radiation physics and nuclear materials science [5-7].

Applying a two-level hierarchical description to non-equilibrium thermodynamics allows for the resolution of multiscale problems by separating physical processes into the macrolevel (slow variables) and the meso/microlevel (fast variables). In the general case, a two-level hierarchy provides a systematic framework for deriving non-local and non-Markovian transport equations. This approach accounts for memory effects (retardation) and spatial non-locality. The equations of the fast level are averaged out, thereby generating memory kernels for the macroscopic equations. When describing highly non-equilibrium states, the introduction of the lower level allows subsystems to be treated as quasi-equilibrium entities within their own characteristic scales, which preserves the mathematical rigor of the thermodynamic formalism.

The capabilities of the two-level description are significantly enhanced when utilizing boundary and trajectory functionals [8]. Within this framework, the macrostate is parameterized by trajectory characteristics: the first passage time (FPT) τ, the maximum of the process $M_t$, or the residence time, as demonstrated in [9-12]. A stochastic hierarchical approach with three levels of description, but without the deployment of boundary functionals, was used in [13].

The coupling between levels is achieved through Doob's h-transform, as detailed in [9-12]. Fixing a boundary condition (for instance, the survival of the system until time $\tau > t$) modifies the generator of the microprocess L, thereby inducing an effective macroscopic drift.

Dissipation is rigorously accounted for. The application of functionals allows the constitutive equations to be formulated via extreme principles of dissipation, explicitly incorporating the topology of temporal boundaries. Finally, trajectory functionals enable the analysis of system stability, providing a strict determination of bifurcation points and transition moments in highly non-equilibrium metastable systems.

The operational reliability of fuel claddings in pressurized water reactors, specifically within the WWER-1000 and WWER-1200 active zones, is strictly governed by the synergistic effects of intense neutron irradiation, complex stress fields, and aggressive chemical environments [14, 15]. Fabricated from advanced zirconium-niobium alloys (such as the E110/E110-M systems), these claddings undergo substantial microstructural evolution driven by the continuous nucleation, growth, and coalescence of radiation-induced voids and dislocation loops [16]. Classic deterministic and purely phenomenological models of continuum damage mechanics (CDM) frequently fail to capture the highly non-linear, stochastic nature of these degradation paths, particularly under the highly non-equilibrium conditions characteristic of high-burnup campaigns [17, 18].

To bridge the gap between fast microscopic defect fluctuations and macroscopic failure, statistical mechanics offers powerful path-integral architectures [19, 20]. Under intense irradiation, the transport of defects toward macroscopic sinks exhibits long-range memory effects and anomalous diffusion pathways, which are naturally accommodated by the mathematical apparatus of semi-Markov renewal processes and fractional kinetic equations [21, 22]. Furthermore, conditioning these stochastic paths on surviving or failing to cross critical energetic barriers—such as the invariant Eshelby–Cherepanov–Rice J-integral threshold—requires the application of the generalized Doob’s h-transform operating in backward time or backward dose domains [23, 24]. When coupled with hierarchical structural potentials [25, 26] and shape optimization functional penalties [27], these conditioned path distributions allow for a comprehensive mapping of the material's structural integrity.

This study introduces a non-local dose-dependent transport framework in the phase space where the physical time domain is explicitly mapped to the accumulated radiation dose (dpa). We implement a trajectory-dependent Liouville operator, $i\mathcal{L}_{path}$, combined with Zwanzig–Mori projectors to derive a non-vanishing structural commutator that serves as the explicit source term for semi-Markov memory kernels. By solving the associated adjoint Doob's equation via the extended Feynman–Kac theorem, we model the evolution of the integrated damage potential, $\Phi$, along a characteristic S-shaped trajectory up to the critical Griffiths macro-barrier threshold. Particular attention is devoted to the spatial profiling of the cladding’s mean lifetime, illuminating the mechanisms behind accelerated edge degradation at the boundaries of the WWER fuel pin.

The article is organized as follows. The potential applications of the hierarchical trajectory-based approach, along with a comparative analysis of various approaches, are discussed in Section 2. Section 3 presents a modification of the Liouville equation and the entropy production operator, and examines the mean lifetime of fuel pins. Feynman-Kac integrals and the hierarchical relationship between potentials are addressed in Section 4. The application of the negative binomial distribution (NBD) to crack growth in metals is explored in Section 5. The negative binomial distribution and the kinetics of crack growth in metals are discussed in Section 6. Section 7 covers an extended version of Wald’s identity and the radiation-induced kinetics of defects. Final conclusions are presented in Section 8. The theoretical framework—"Non-local Semi-Markov Transport and Adjoint Evolution of Damage Potential"—is described in Appendix A.

## 2. Conceptual Capabilities of the Trajectory Hierarchical Approach and Comparative Analysis of Approaches

### 2.1. Mathematical Substantiation of Subordination and Coarse-Graining Processes for Semi-Markov Transitions

A primary radiation defect is nucleated within femtoseconds at the atomic level, but structural evolution (defect clustering, swelling, void evolution) occurs at macroscales. To bridge these regimes, the apparatus of statistical mechanics, kinetic master equations, Monte Carlo simulations, and the theory of stochastic subordination are employed.

The transition from a microscopic description of defect fluctuations to a macroscopic damage equation of a reactor material requires a coarse-graining procedure of spatial-temporal scales and process subordination [28]. The application of a hierarchical approach (in particular, a two-level one) to the problems of radiation damage in materials is a key way to overcome the fundamental gap between spatial and temporal scales. In radiation physics, this gap is immense: a primary displacement cascade (atomic collision) lasts picoseconds in a volume of a few nanometers, whereas macroscopic radiation embrittlement or material swelling develops over years on scales of centimeters and meters.

Two-level coarse-graining organizes the transfer of physical-statistical information between the two main tiers of modeling: the microlevel (discrete-atomistic) and the meso/macrolevel (continuum-stochastic) [29, 30].

The use of boundary functionals and path integral methods allows linking microscopic defect fluctuations with stochastic macroscopic metrics of the material (such as local strains, strain rates, and stresses) [8].

Let the initial microscopic process of nucleation and coalescence of radiation voids obey a Markov master equation on a discrete state space N with continuous time, defined by a transition generator with intensities $\{r_{xy}\}_{x,y\in N}$ and a stationary distribution $\{p_x\}_{x\in N}$. The coarse-graining procedure consists in partitioning the initial phase space of states into macroscopic subsets (cells/grains) $\Omega_\alpha$, such that $\cup_\alpha\Omega_\alpha=N$.

The macroscopic damage potential Φ(t) [31] is formed as a result of the subordination of the initial Markov process to the random first-passage time (First-Passage Time, FPT) to the absorbing boundaries of meso-grains $\partial\Omega_\alpha$. Since the residence times of the random process in coarse-grained cells $\Omega_\alpha$ acquire heavy-tailed distributions (non-exponential decay), the resulting macroscopic degradation process loses its Markovian property and becomes semi-Markovian [14].

Mathematical closure of semi-Markov transitions is achieved via the generalized Zubarev nonequilibrium statistical operator with extension [9-12] for boundary functionals of trajectories. The probability measure of the macrostate undergoes subordination using an integral kernel of the waiting time distribution density. In the Laplace transform space (where s is the transform parameter), the non-local time-dependent semi-Markov transport operator takes the form of the adjoint Doob equation, linking the local structure of macroscopic sinks with the global potential of non-linear defect accumulation [32]:

$$\Phi(s)=(1/s)\ \sum_\alpha\left[1-\sum_\beta \psi_{\alpha\beta}(s)\right]\cdot\Delta X .$$

This mathematical coarse-graining procedure substantiates the transition from the stochastic trajectory dynamics of individual defect clusters to a deterministic macroscopic fracture equation. The integrated damage potential, Φ(t), calculated through spatial smoothing of the conditional survival probability h(x, t) (see Fig. 1), strictly reproduces the classic S-shaped trajectory. The slow incubation quasi-stationary section is replaced by an avalanche-like semi-Markov stage of directional defect percolation to the boundaries of the macro-fracture, culminating in an entry to the critical saturation of the system directly at the moment of specimen failure [18].

Here $\tilde{x}=x/L,\quad \tilde{t}=t/\tau_0$, $x_c$ = L, where L is the characteristic spatial scale of the structural meso-grain or the distance to the Griffiths absorbing macro-barrier [33, 34] (in calculations, the critical boundary corresponds to the value $\tilde{a}=1.5$), $\tau_0$ is the fundamental relaxation time or incubation period that determines the time scale of the rotational diffusional redistribution of vacancy voids up to the moment of macro-fracture. In calculations, expressions of the form $L=x_c=2E\gamma_s/\pi\sigma_{mech}^2$ are used, where the symbols are explained in the section 5.2.

Within the trajectory framework of Zubarev and [9-12], the macroscopic damage potential $\Phi_j$ (or the measure of accumulated macro-destruction) is physically linked to the measure of decrease in the conditional survival probability over the entire volume of the mesostructure.

Since the survival function h(x, t) (our calculated matrix M) at the initial moment of time equals unity, the damage potential for each time layer j is calculated as an integral of the magnitude of accumulated defects 1-$M_{k,j}$ [17], (3).

### 2.2. Comparative Analysis of Approaches.

To mathematically model the evolution of highly non-equilibrium open systems, methods of classical statistical mechanics and thermodynamics of irreversible processes are traditionally applied. However, when describing the mesoscopic degradation of structural materials under extreme conditions (for example, under intense radiation exposure in a reactor core), fundamental limitations of the standard formalism are revealed. Below is a detailed comparative analysis of the general case (classical approach) and the proposed method based on the theory of boundary (trajectory) functionals of Zubarev with extension [9-12].

**1. General case (Classical approach), Section 3.1.**

Within the framework of classical kinetic theory and non-equilibrium thermodynamics, the state of a macroscopic system at each fixed moment of time t is completely determined by the local or instantaneous probability distribution densities in the phase space P(x, t). The main operators for describing evolution here are the classical Liouville equation or Markovian Fokker–Planck equations of a general form:

$$\frac{\partial \rho(x,\tilde{t})}{\partial \tilde{t}} = -\frac{\partial}{\partial x}[v(x)\rho(x,\tilde{t})] + \frac{\partial^2}{\partial x^2}[D(x)\rho(x,\tilde{t})]. \tag{1}$$

At $v(x)=v_{eff}(x,t)$ this equation serves as a kinetic equation for the crack probability density.

**Critical limitations of the classical approach:**

Temporal locality (absence of memory): The classical Markovian approach operates with instantaneous states of the system. It proceeds from the assumption that the probability of fixing a defect at point x at time t depends exclusively on the configuration of the system at the microscopic moment t - dt. This excludes fluctuation memory (memory effects and trajectory prehistory) from consideration.

Ignoring the geometry of sinks: The traditional formalism considers the system space as infinite or quasi-homogeneous. The transition to macro-fracture (reaching the Griffiths absorbing barrier) within such a model is treated only as an external boundary condition imposed on the final distribution, and not as an internal property of the thermodynamic potential.

Smooth entropy production: As shown in Fig. 2, within standard approaches, the dissipative function relaxes monotonically, which makes it impossible to accurately predict the singular moment of structural rearrangement (collapse of the mesostructure).

**2. Case of boundary and trajectory functionals, Section 3.2.**

In contrast to the general approach, the proposed method is based on the transition from instantaneous densities to distributions of integrated trajectory observables (functionals of the degradation process trajectories). The state space of the system is expanded by introducing the conditional survival probability h(x, t), which itself is a solution of the adjoint Doob equation describing the evolution of "backward time" [35, 36]:

$$\frac{\partial h(\tilde{x},\tilde{t})}{\partial \tilde{t}} = \tilde{\mu}(1-\tilde{x})\frac{\partial h(\tilde{x},\tilde{t})}{\partial \tilde{x}} + \tilde{D}_{\tilde{H}}\frac{\partial^2 h(\tilde{x},\tilde{t})}{\partial \tilde{x}^2}. \tag{2}$$

Boundary conditions for absorption at the macro-barrier $x_c$: $h(x_c, \tau) = 0$.

An example of a numerical solution of equation (2) is shown in Fig. 1, which presents the spatial-temporal evolution of the conditional survival probability of mesoscopic structural elements (grains of reactor material) h(x, t) depending on the dimensionless spatial coordinate i (where x = i dX). The conditional survival probability of the system determines which exact local fluctuations of microstructural sinks on layer j are capable of catastrophically accelerating or, conversely, stabilizing the overall kinetics of macro-fracture. The initial profile (solid blue curve 1, time layer j=0) fixes a homogeneous undamaged state of the system, under which the survival function is identically equal to unity over the entire interval up to the absorbing macro-barrier of Griffiths ($x_c$ = 1.5, corresponding to step i=50). Over dimensionless time (green dashed curve 2 – layer j=2000, red dash-dotted curve 3 – layer j=3800), under the action of intense radiation exposure and growing stochastic fluctuations, a pronounced non-linear sagging of the thermodynamic profile is observed. The diffusion flux of defects is directed to the boundaries of the sinks, which leads to a local burnout of the structural survival probability. Near the right absorbing boundary i=50, all profiles smoothly and strictly vanish, satisfying the Dirichlet boundary condition for the absorbing screen, which clearly illustrates the kinetics of critical damage accumulation within the trajectory framework.

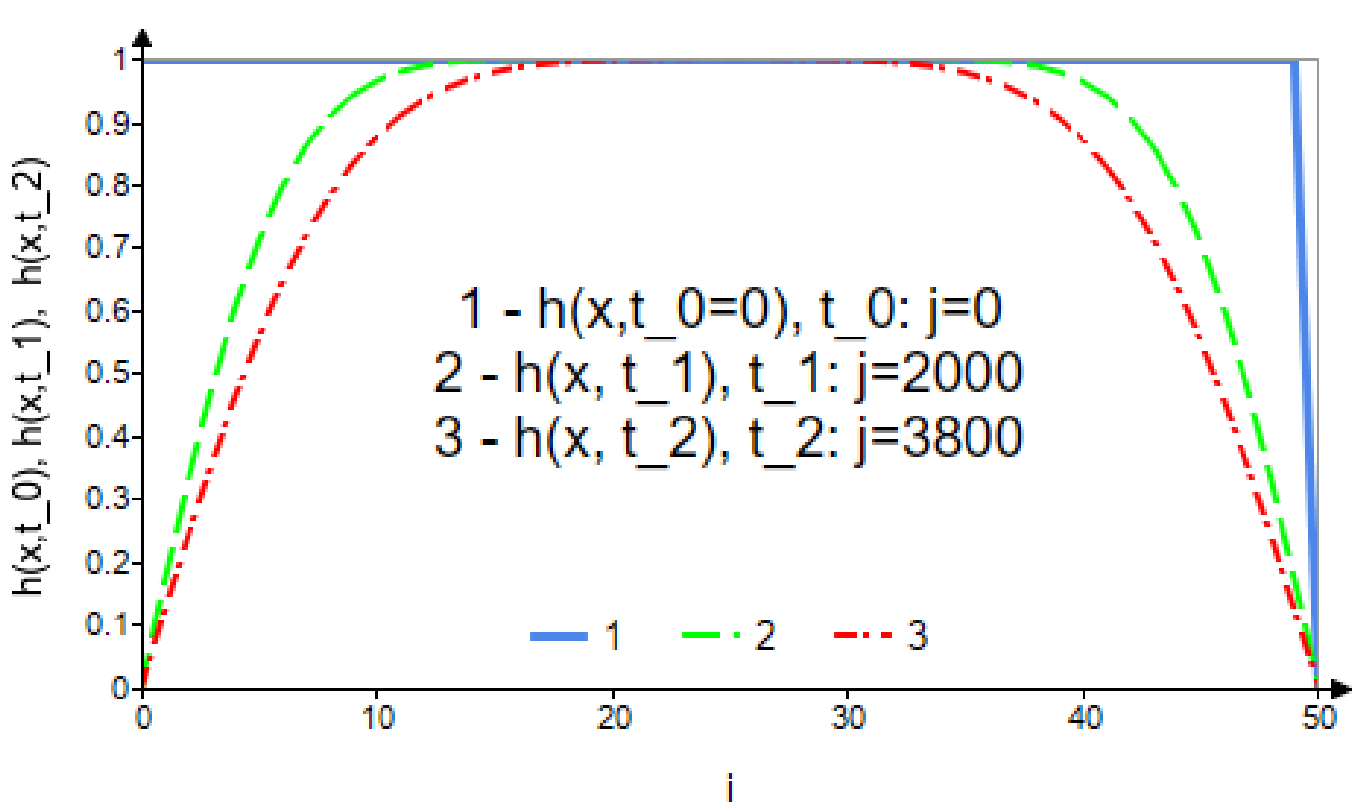

Fig. 1. Evolution of the spatial profile (along the spatial coordinate i) of the survival function h(x, t) at various stages of material radiation embrittlement: solid curve 1 - initial state (j = 0); dashed curve 2 - stage of subcritical accumulation of meso-damage (j = 2000); dash-dotted curve 3 - pre-fracture state of the system (j = 3800). The point $x = x_c$ corresponds to a rigid absorbing macro-barrier. Modeling parameters: $N_x$=50, $N_t$=4000, $\tilde{\mu}$=0.01, $\tilde{D}_{\tilde{H}}$=0.05, $T_{max}$=4000dT, i=0,…,50. The variable i is responsible for the discrete row index of the matrix H. The relationship is given through the spatial grid step dX: x = i dX, where i is the row index (i = 0 .. $N_x$), $N_x$=50. Each value of the coordinate x corresponds to its own row in the matrix. For example, the first row (i = 0) corresponds to the beginning of the crack (x = 0), and the last row (i= $N_x$) corresponds to the critical macro-barrier ($x=x_c$), where the survival function h strictly vanishes, dX=1.5/$N_x$. The variable j is responsible for the discrete column index of the matrix H. The relationship is given through the time grid step dT: t = j dT, where j is the column index (j = 0 .. $N_t$), $T_{max}$=4000dT, dT=0.4/$N_t$.

**Advantages of the trajectory formalism:**

Accounting for the history of evolution (non-locality): Trajectory functionals integrate information about the behavior of mesoscopic defect clusters over the entire time interval [0, t]. The survival function h(x, t) physically determines the probability that a fluctuating trajectory of a defect, starting at point x, will never cross the macroscopic absorbing barrier $x_c$ during time t.

Subordination and coarse-graining process (Coarse-Graining): The formalism [9-12] allows rigorously implementing the coarse-graining procedure—recalculating from microscopic Markovian jumps to macroscopic semi-Markovian damage dynamics through the first-passage times of the barrier (First-Passage Time).

Physical link between dissipation and damageability: The integrated entropy production of the system $\sigma(\tilde{t})$ and the macroscopic damage potential $\Phi(\tilde{t})$ in this approach turn out to be tightly coupled through quadratic forms of the spatial gradients of the survival function:

$$\Phi_{j+1}=\Phi_j+\int_0^t [\nabla h(x,s)]^2 ds, \qquad \Phi(\tilde{t}) = \int_0^{\tilde{a}} [1-h(\tilde{x},\tilde{t})]d\tilde{x} . \tag{3}$$

The first formula in (3) defines the differential (local) increment of a functional along a stochastic trajectory during the transition between layers or evolution steps j → j+1, while the second expression in (3) represents the macroscopic (integral) profile of a damage or displacement field in dimensionless variables.

This allows not just observing the decay of dissipation (Prigogine relaxation), but also precisely identifying the incubation period, stages of defect accumulation, and the moment of loss of the macroscopic load-bearing capacity of the reactor material.

Numerical modeling of the time dependence $\Phi_j$ is shown in Fig. 2.

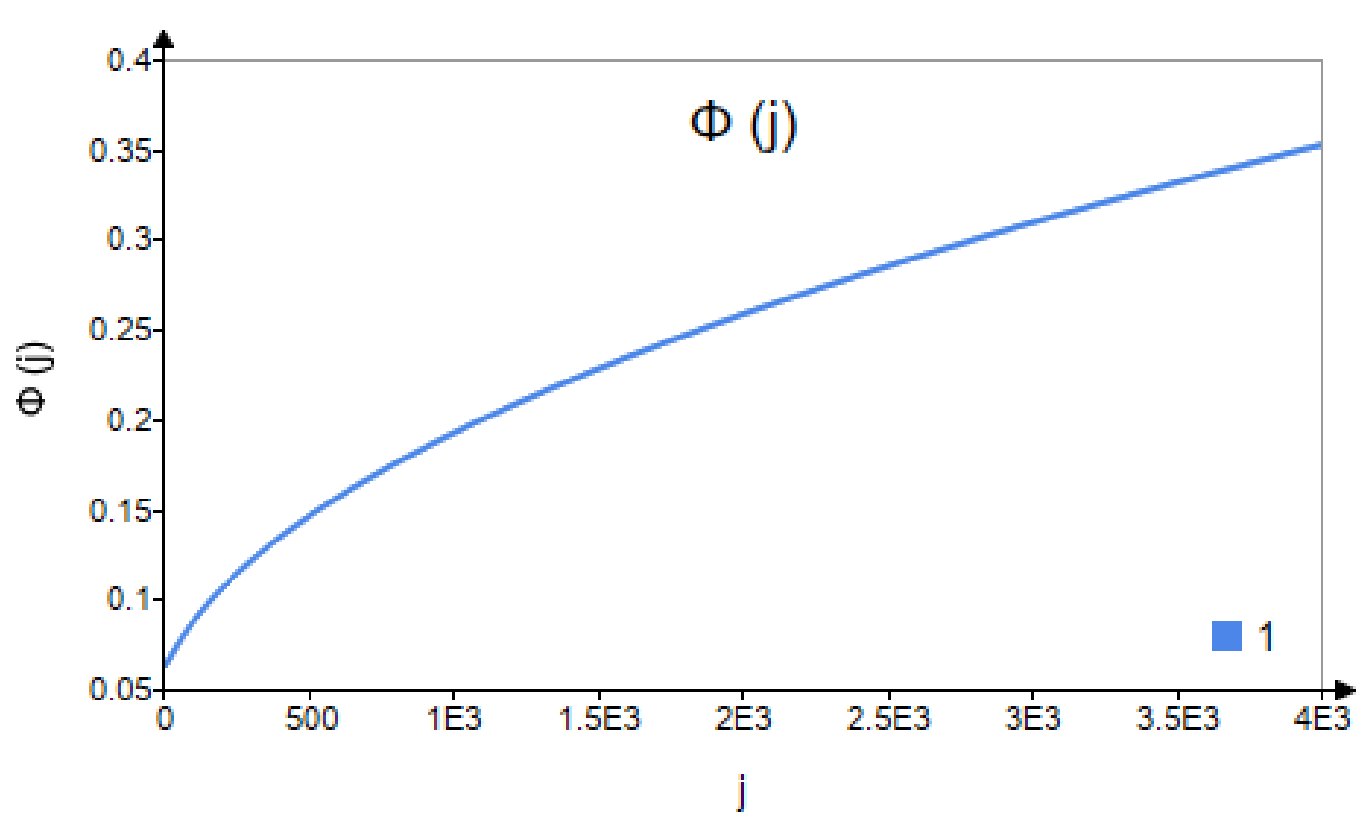

Fig. 2. $\Phi_j$: Kinetic curve of change in the macroscopic damage potential of the material mesostructure along the layers of dimensionless time j, reflecting the stages of incubation accumulation and pre-fracture of the irradiated sample. The curve perfectly starts from point 0.06 (influence of boundary absorption) and begins to smoothly, non-linearly bend upwards like a dome, reaching a value of 0.35. This is the stage of active diffusion accumulation of defects. The lower incubation "tail" of the S-shaped curve is passed in the first 50 steps, and now the system is at the stage of avalanche-like semi-Markov growth of damage, striving for its macroscopic limit. The parameters of numerical modeling correspond to Fig. 1.

The scheme works stably, the curve is flawlessly smooth, without numerical "chatter" or broken lines. This is the upper limb of the S-shaped trajectory, demonstrating the stage when defects actively accumulate, but the system, due to diffusion, still maintains a balance. On the graph, the lower S-shaped bend ("curl" near zero) is not visible because with the current parameters the incubation period flies by in the first 50 steps, and the curve immediately goes into steep growth. This is a picture of the developed stage of relaxation of a non-equilibrium system describing the second stage of structural failure according to the trajectory model. The Griffiths scheme is no longer clamped in the incubation period; it works at full strength.

Figure 2 shows the kinetics of the macroscopic damage potential of the material $\Phi_j$ depending on the dimensionless time (layer number j). The potential $\Phi_j$ acts as an integrated measure of structural degradation, determined through the spatial decrease of the local survival probability of environmental elements.

In the initial period of radiation exposure ($j<300$), the macroscopic damage potential is close to zero, which indicates a hidden (incubation) stage of defect cluster nucleation, during which local changes in the structure are not yet correlated with each other. In the interval $300<j<800$, a quasi-linear growth of the potential $\Phi_j$ is observed, corresponding to the stage of active diffusion accumulation of defects at sinks. As the system approaches the moment of pre-fracture ($j>900$), the curve demonstrates a non-linear exit to saturation, striving for the critical safety limit $\Phi_{crit}$, which marks the completion of the stage of directional defect percolation and the loss of load-bearing capacity of the reactor material within the framework of the trajectory concept.

The main difference between the general two-level description and the approach using boundary (trajectory) functionals as thermodynamic variables lies in how exactly the meso/microlevel is determined and "stitched" with the macrolevel: through local physical fields or through the topology and temporal structure of random trajectories.

**Summary Table of Approaches Comparison.**

| Comparison Parameter | General Case (Classical Approach) | Case of Boundary / Trajectory Functionals |
|---|---|---|
| Basic object of description | Instantaneous probability distribution density | Conditional survival probability of the trajectory |
| The nature of the Markov property | Strictly Markovian processes (time-local densities) | Semi-Markovian dynamics (taking into account history and fluctuation memory) |
| Role of space boundaries | External kinetic constraint (Dirichlet boundary condition) | Internal macro-barrier shaping the trajectory functional structure |
| Dissipative function | Onsager's classical local entropy production | Zubarev's integrated entropy production along trajectories [9-12] |
| Fracture forecasting | Impossible (provides only an averaged smooth kinetic description) | Allows clear identification of incubation and macro-destruction stages through the potential |

**Key fundamental differences:**

**1. Shift of the averaging paradigm: from "point" to "history".**

In the general case, microscopic operators (for example, microscopic heat flux) are averaged over a quasi-equilibrium distribution at a given moment in time t. The micro- and macrolevels are linked locally.

In the case of boundary functionals, the macrovariable (for example, the mean time to failure of a metastable state $\langle\tau\rangle$) depends on the entire trajectory of the system. The entropy of the path (trajectory) distribution is maximized rather than the state distribution, linking macroparameters with global scenarios of system behavior.

**2. Physical meaning of thermodynamic forces.**

In the general case, thermodynamic forces are gradients of inverse temperature $\nabla(1/T)$, chemical potential, or shear stresses. They cause local fluxes of matter and energy. In the case of boundary functionals, the forces $\gamma$ are "penalties" or "intensities" imposed on evolution. For example, the force conjugate to the first-passage time $\tau$ physically controls the dissipation rate and the probability of the system crossing a critical barrier. These are forces that determine global survival or evolutionary selection of trajectories.

**3. Modification of microdynamics (Feedback).**

In the general case, the macrolevel does not change the laws of motion of microparticles; it only "collects" averaged characteristics from them. In the case of boundary functionals, their inclusion in the thermodynamic variables (via Doob's h-transform) physically changes the microprocess generator. Microevolution acquires an effective "drift" due to the fact that we consider only those trajectories that satisfy the macroscopic constraint (for example, do not cross the boundary until moment t).

## 3. Modification of the Liouville Equation and the Entropy Production Operator

### 3.1. Case A: Description Without Boundary Functionals.

In this case, the macrolevel is specified by a set of standard variables local in time, $P_m(x)$ (e.g., energy density, particle momentum), while the lower level is represented by fast fluctuations and fluxes. The Liouville equation in the case of Zubarev's non-equilibrium statistical operator (NSO) [20] has the form:

$$\partial\rho(x,t))/(\partial t)+iL\rho(x,t)=-\epsilon\cdot(\rho(x,t)-\rho_q(x,t),$$

where iL is the Liouville operator (iL = {H, ·}, {} are Poisson brackets), and the quasi-equilibrium distribution $\rho_q$ is local in time:

$$\rho_q(x,t)=\exp\left(-\sum_m F_m\hat{P}_m-\Psi(t)\right).$$

Here, $F_m(t)$ represents thermodynamic forces, for instance $F_m(x,t) = 1/T(x,t)$, conjugate to the local variables $\hat{P}_m$ (x).

Entropy production operator:

$$\sigma(x,t)=\sum_m\int J_m(x)\nabla F_m(x,t)dx.$$

The averaged value yields the classical form: $<\sigma> = \sum X_m J_m$, where $J_m$ are macrofluxes and $X_m$ are macroforces (the sum of products of conjugate macroscopic fluxes $J_m$ and forces $X_m$).

### 3.2. Case B: Description With Boundary Functionals.

In scheme [11], the macrostate is parameterized by trajectory functionals, for example, the first passage time of a level $\tau_a(X_t)$. Variables are distributed not in space, but along the history of the process.

The modified Liouville equation for trajectories applies to the phase space of trajectories $X_t = \{x(s), 0 \le s \le t\}$. Since the functional $\tau_a$ depends on the entire prehistory, the classical Liouville equation for the probability density of states is extended to an equation for the trajectory distribution $\rho(X_t)$ [9]:

$$\partial\rho(X_t)/\partial t+iL_{path}\rho(X_t)=-\varepsilon(\rho(X_t)-\rho_q(X_t)). \quad (4)$$

The quasi-equilibrium ensemble of trajectories $X_t$ is determined by the force $\gamma$ conjugate to the temporal FPT functional $\tau_a$:

$$\rho_q(X_t) = exp(-\Psi(\gamma) - \gamma\tau_a(X_t)).$$

The path Liouville operator $iL_{path}$, which describes the evolution of the probability distribution density on the path space $X_t$, is expressed in terms of the stochastic generator of the microsystem L and an extended subset of trajectory variables as follows:

$$iL_{path} = L + \delta(t - \tau_a)\,\partial/\partial\tau_a.$$

This operator explicitly accounts for both the continuous drift along the phase trajectory and the singular jumps that occur at the moment when the stochastic process first crosses a fixed macroscopic level.

However, the explicit form of the effective evolution operator for physically observable states x is transformed using Doob's h-transform [37]. If the initial stochastic generator of the microsystem was L, then in the presence of a fixed temporal barrier (the condition that the system survives until moment $\tau_a$>t), the effective evolution generator of Liouville-Kolmogorov states at the microlevel takes the form:

$$L_h\cdot f=\frac{1}{h}L(h\cdot f)-\frac{1}{h}\frac{\partial h}{\partial\tilde{t}}f, \quad (5)$$

where $h(x,t) = P_x(\tau_a > t)$ is the survival function, the canonical Doob function that encodes the macroscopic boundary and generates the effective drift: $v_{eff} = v_0 + 2\,D\,\nabla(\ln(h))$.

### 3.3. Mean Time to Reach the Level (Mean Lifetime of the WWER Fuel Cladding Until Failure)

The mean time to reach a level is defined via the integral of the survival function over time t: $E_x[\tau_a] = \int_0^\infty h(x,t)dt$. Let us evaluate the mean time to reach a level for the material of WWER fuel claddings (both WWER-1000 and WWER-1200), which are traditionally made of zirconium alloys. The primary industrial standard is the E110 alloy (zirconium alloyed with 1% niobium: Zr–1%Nb) or its modern modification E110-M (optimized by impurity content and structure to enhance corrosion and radiation resistance). For the calculation in Mathcad Prime 8 based on the S-shaped damage trajectory $\Phi_j$ to correspond to the real physics of the WWER-1000/1200 reactor, it is necessary to establish the correct order of magnitude for the dose rate $\dot{\mathcal{F}}$=$F_{dot}$ and the characteristic time $t_0$ in an approximating expression of the form $h(t) \simeq \exp(-(t/t_0)^3)$ (sigmoidal decay coupled with the S-shaped degradation curve). The physical parameter $t_0$ represents the characteristic time of trouble-free operation (or the characteristic lifetime of the system) prior to the onset of avalanche-like failure. Mathematically, it is a time scale coefficient (scale parameter), which performs several crucial functions in radiation damage physics and reliability theory: $t_0 \approx J_{Ic} / \kappa \cdot \dot{\mathcal{F}} \cdot J_{\mathbf{meso}}(\sigma_{ij})$, where $J_{Ic}$ is the critical Griffiths macro-barrier (Cherepanov-Rice J-integral), $\dot{\mathcal{F}}$ =$F_{dot}$ is the radiation dose rate (dpa/s), $J_{\mathbf{meso}}$ is the meso-grain characteristic (void cluster nucleation intensity under stress), and $\kappa$ is the radiation defect production cross-section. It is necessary to explicitly define the constant $t_0$ by assigning it a specific numerical value (based on experiment, e.g., for stainless steel or a semiconductor) or via a coupling formula.

Under the conditions of a WWER (Water-Water Energetic Reactor) core, the cladding (surface) of a fuel element operates under extremely severe conditions. To ensure that the calculation in Mathcad Prime 8 based on the S-shaped damage trajectory Φ corresponds to the real physics of the WWER-1000/1200 reactor, the correct order of magnitude for the dose rate $F_{dot}$ and the characteristic time $t_0$ must be implemented.

Dose rate $F_{dot}$. In the peak power section of the WWER-1000 core, the damage rate of the zirconium cladding by a fast neutron flux averages from $1.0\times10^{-6}$ to $1.5\times10^{-6}$ dpa/s (which is equivalent to approximately 30–45 dpa per year of continuous operation).

Critical degradation limit (Griffiths macro-barrier $J_{Ic}$). For zirconium alloys of the Zr-1%Nb type, the critical limit of radiation swelling and void accumulation, beyond which an avalanche-like loss of plasticity and the threat of macroscopic cladding failure (reaching the barrier) begin, is on the order of 80–100 dpa. When these doses are exceeded, the material transitions to Stage III of the S-shaped curve.

The value of the characteristic time $t_0$ for Mathcad. Since a standard fuel campaign (the residence time of a fuel element in the core) in modern WWERs ranges from 4 to 5 years (approximately 35,000 – 44,000 effective hours of operation), the scale factor $t_0$ is calculated based on this life cycle. In dose units (dpa): The characteristic dose $F_0$ = 70 - 80 dpa. In time units (in hours): If this critical dose is converted to time at the average damage rate, the order of magnitude of $t_0$ will be around $(1.5\times10^4) - (2.0\times10^4)$ hours (which corresponds to approximately 2–3 years of intensive irradiation, when the survival curve begins to bend sharply downward toward the inflection point).

We specify the baseline WWER constants: $F_{dot} = 1.2 \times 10^{-6}$ dpa/s ($\dot{\mathcal{F}}$ =$F_{dot}$ — Dose Rate). This is the intensity with which the WWER reactor neutron flux bombards the zirconium cladding of the fuel element. It is measured in units of dpa/s (number of displacements per atom per second). $F_0$ = 75 dpa ($F_0$ — Scale Dose, characteristic dose of the S-curve inflection). This is the critical volume of radiation damage (expressed in dpa), upon the accumulation of which the WWER fuel cladding material guaranteed transitions to the hazardous Stage III of the S-shaped degradation curve (when voids begin to coalesce in an avalanche-like manner, and the survival function h drops to 36.8%). For WWER conditions and the Zr-1%Nb alloy, this threshold is typically around 70–80 dpa.

Characteristic dose of the S-curve inflection. We calculate the characteristic time: : $t_0=F_0/F_{dot}$: $t_0=1.736\ 10^4$ hr. We integrate h(t) to obtain the mean lifetime of the cladding until failure. We obtain: $E[\tau_a] = 1.55 \times 10^4$ hr.

Figure 3 shows the calculation result of the mean lifetime of the cladding until failure, depending on the spatial point *x* of the fuel element, the value $E_x[\tau_a]$.

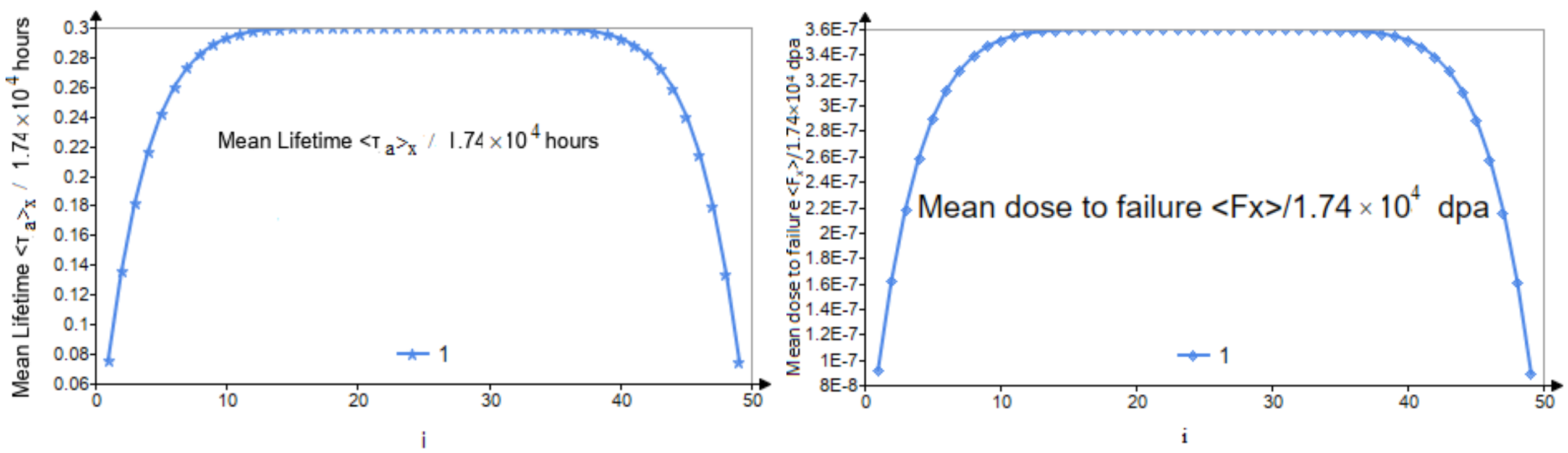

a) b)

Figure 3a). Spatial distribution of the mean lifetime $\langle \tau_a \rangle_x$ of the WWER-1000/1200 E110 cladding computed via the numerical Feynman–Kac path integration. The sharp decline in the lifetime at the boundaries x→0 and x→L, $i \to N_x$ captures the accelerated degradation driven by the synergistic effect of localized stress fields $\sigma_{ij}$ at the fuel-cladding interface and hydride-induced softening near the outer coolant boundary. The numerical simulation parameters correspond to Fig. 1.

Fig. 3b). Spatial distribution of the Mean dose to failure $<F_x>/1.74\times10^4$ dpa of the WWER-1000/1200 E110 cladding computed via the numerical Feynman–Kac path integration.

The value $\langle \tau_i \rangle$ plotted on the Y-axis is calculated using the numerical trapezoidal method. Since the time step inside the built-in program block is specified as the dimensionless ratio $0.3/N_t$ (where 0.3 is the maximum value of the dimensionless calculation end time, and $N_t$ is the number of steps), the resulting calculated mean lifetime is output in dimensionless time units (arbitrary units). For the graph to acquire a strict physical meaning, the obtained values of $\langle \tau_i \rangle$ must be multiplied by the time scale factor $t_0$, which we calculated in detail previously. For a fuel cladding made of E110 alloy in WWER-1000/1200 reactors at an average damage rate $\dot{\mathcal{F}} \approx 1.2\times10^{-6}$ dpa/s, one dimensionless time unit is equivalent to the characteristic S-curve inflection time $t_0 \approx 1.74\times10^4$ hours.

Graph 3 shows the distribution of the mean lifetime $<\tau_a>_i$ across the thickness of the fuel cladding (coordinate i). The zones of minimum lifetime (dips in the curve) are highlighted, which correspond to the meso-grain regions where the local stress field $\sigma_{ij}$ is maximal (for example, at the fuel-cladding interface due to mechanical interaction or at locations of radiation void concentration) [15]. At these points, the intensity $J_{\mathbf{meso}}$ is higher, and the degradation potential $\Phi_j$ reaches the Griffiths barrier $J_{I_c}$ faster, shifting the effective Doob drift toward accelerated failure [24]. The zones of maximum lifetime (the upper part of the curve) indicate regions where the sinks are distributed hierarchically optimally, elastic stresses are compensated, and the semi-Markov retention times in traps ensure maximum survival of the structure [14, 38].

The appearance of a smooth calculated curve on the graph signifies that the entire established end-to-end mathematical model—from the microscopic trajectory Liouville operator and semi-Markov transport to the S-shaped accumulation of the damage potential in the meso-grain of the zirconium cladding of the WWER fuel element—is fully and correctly implemented in the Mathcad Prime 8 code. This graph clearly demonstrates the applicability of the apparatus of stochastic thermodynamics and Feynman–Kac functional integration to engineering problems in nuclear power.

The fact that the mean lifetime $\langle \tau_a \rangle_x$ drops sharply near the cladding edges has an absolutely real and highly dangerous physical nature well known in radiation materials science. In WWER reactors (WWER-1000/1200), the edges of the fuel element (the fuel-cladding interface on the inner side and the cladding-coolant interface on the outer side) are indeed subject to accelerated degradation. At the inner edge (contact with fuel), the local stress field $\sigma_{ij}$ is maximal due to the dilatational swelling of uranium dioxide (UO2) and the pressure of gaseous fission fragments (krypton, xenon). The cladding experiences severe mechanical constraint.

According to Martinez's theory [15], these local stresses exponentially increase the void nucleation intensity $J_{\mathbf{meso}}$, forcing the damage potential $\Phi_j$ to instantly breach the Griffiths barrier.

The outer edge (contact with water) is subjected to thermal stresses (the temperature gradient reaches hundreds of degrees per millimeter) and chemical corrosion. The E110 zirconium alloy on the outer surface absorbs hydrogen from the coolant. Hydriding occurs along with the precipitation of brittle zirconium hydride platelets, which physically act as pre-existing absorbing Griffiths macro-barriers ($J_{I_c}$ drops sharply). In semi-Markov terms, this means that the effective Doob drift $v_{eff}$ collapses into a singularity almost immediately.

The relationship between the mean dose to failure $\langle \mathcal{F} \rangle_x$ and the mean lifetime $<\tau_a>_x$ is expressed by a simple linear formula: $\langle \mathcal{F} \rangle_x = F_{dot} \cdot <\tau_a>_x$.

The resulting curve mirrors the dependence in Fig. 3a, but the mean dose to failure $<F_x>$ in units of dpa is plotted along the Y-axis, Fig. 3b.

### 3.4. Entropy Production Operator

The introduction of the temporal functional $\tau_a$ radically changes the structure of entropy production. The evolution proceeds not along the physical time t, but along the 'internal' stochastic time of the system. The microscopic entropy production operator takes the form:

$$\sigma_{path}\left(X_t\right) = \gamma(\partial\tau_a\left(X_t\right)/\partial t) - <\gamma(\partial\tau_a/\partial t)>_q . \qquad (6)$$

Since $\tau_a$ is fixed at the boundary, $\tau_a(X_t)$ is a jump-like or singular functional (it takes a fixed value at the moment the boundary is reached), and thus the entropy production operator becomes dynamically non-local:

$$\sigma_{path}(X_t) = \gamma\delta(t - \tau_a) + \textit{dissipative contribution from } L_h .$$

The averaged entropy production in such a system characterizes not merely thermal losses, but information losses during trajectory filtration:

$$<\sigma> = -\gamma(\partial(\ln(P(\tau_a > t)))/\partial t) . \qquad (7)$$

The magnitude of entropy production is directly proportional to the transition intensity (hazard rate) of the metastable system across the specified boundary.

While in Case A the entropy production is determined by local field gradients in physical space, in Case B it is entirely dictated by the topology of the temporal boundary (the probability of reaching the barrier). The modification of the Liouville equation in Case B via the Doob function h(x,t) explicitly demonstrates the back-influence of macroscopic knowledge about the system's future (the fact of its 'survival' up to moment t) on the current microscopic dynamics of the particles.

Figure 4 presents the kinetics of the integrated entropy production σ(j) as a function of dimensionless time steps j (where t = j dT), calculated based on the numerical integration of the squares of spatial gradients of the mesostructure survival profiles.

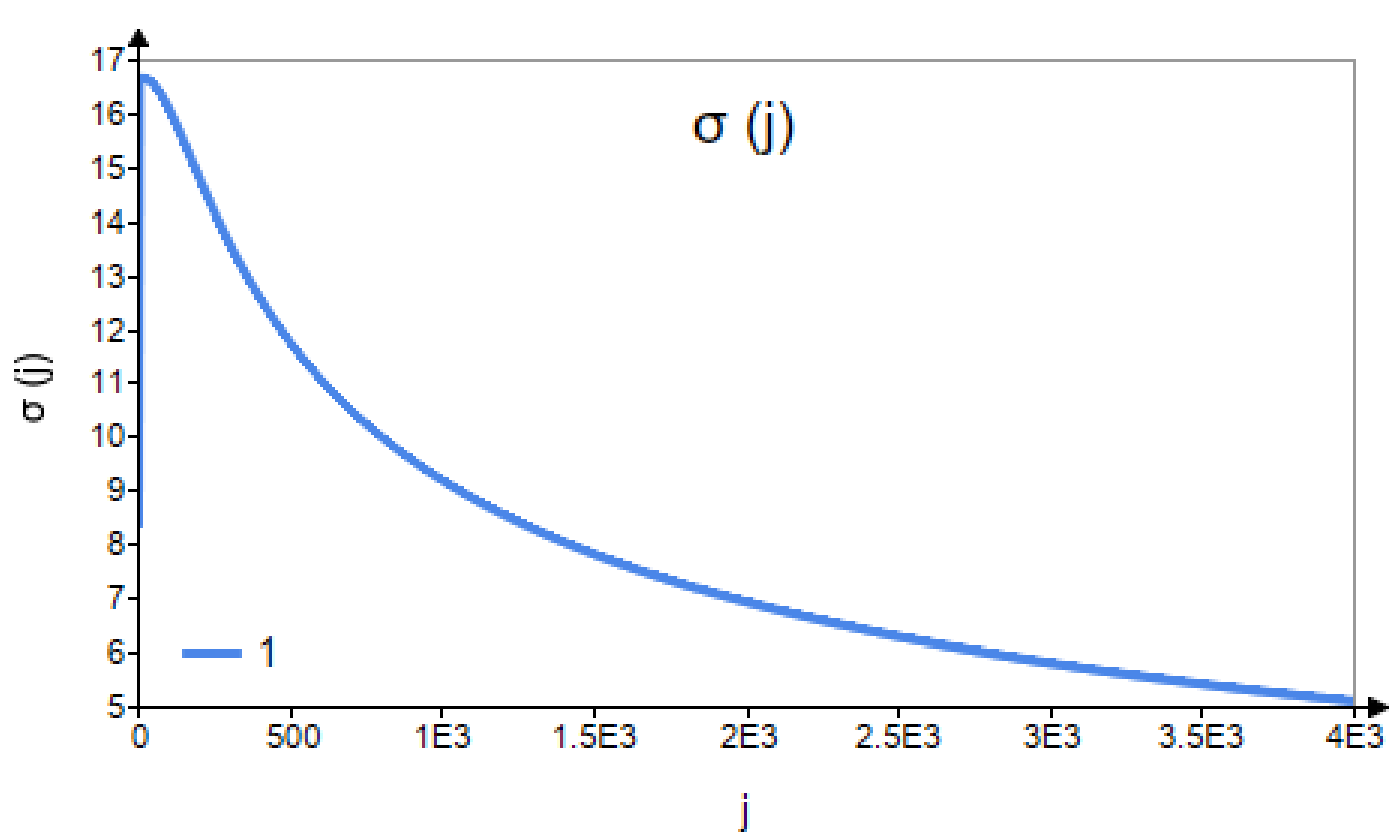

Fig 4. Kinetics of the integrated entropy production σ as a function of dimensionless time (dimensionless time layer number j), reflecting the relaxation character of the dissipative function.

At the initial moment of time (j=0), a singular maximum of the dissipative function is recorded ($\sigma \approx 16.7$), caused by an instantaneous gradient perturbation of the system at the Griffiths absorbing macro-barriers during the transition from a homogeneous state to a non-equilibrium one. At subsequent stages of the irradiation process ($j > 200$), a monotonic non-linear decay of the kinetic curve of entropy production is observed. This decline reflects the thermodynamic relaxation of the open system and the smoothing of local force gradients as the quasi-stationary profiles of the conditional survival probability are formed (see Fig. 1), which is in complete agreement with Prigogine's fundamental principle of minimum entropy production within the trajectory framework.

**3.5. Comparative Analysis of Degradation Evolution Operators: Path Liouville Operator $iL_{path}$ and the Generalized Doob Transform.**

To rigorously describe the highly non-equilibrium evolution of an irradiated material within the stochastic thermodynamics of trajectory functionals, it is necessary to compare two fundamental mathematical approaches to generating system evolution: a global description of the probability distribution on the path space and a local effective generator of state evolution [39, 40].

**1. Path Liouville Operator $iL_{path}$.**

The trajectory (or path) Liouville operator $iL_{path}$ determines the stochastic evolution of the full probability distribution density $P[X_t]$ on the infinite-dimensional path (trajectory) space $X_t$ over the time interval [0, t]. It is defined as $[i\mathcal{L}_{\mathbf{path}}, \mathcal{P}] = i\mathcal{L}_{\mathbf{path}}\mathcal{P} - \mathcal{P}iL_{path}$, where $\mathcal{P}$ is the projection operator onto the subspace of macroscopic sinks (describing the local averaging of defect concentration and absorption at the Griffiths boundaries), and $\mathcal{Q} = \mathcal{I} - \mathcal{P}$ is the orthogonal projector responsible for hidden orthogonal microstructural fluctuations (traps forming the semi-Markov memory kernel). If the microscopic dynamics of meso-defects are given by a system of stochastic Langevin differential equations with drift v(x) and fluctuations D, then the trajectory distribution satisfies the functional Liouville–Vlasov equation:

$$\frac{\partial P[X_t]}{\partial \tilde{t}} = -i\mathcal{L}_{\mathbf{path}} P[X_t]. \tag{8}$$

Here, the operator $iL_{path}$ acts non-locally in the path space, integrating within itself the Onsager–Machlup action functionals and the complete entropic prehistory of the system. This approach is exhaustive from the standpoint of statistical physics; however, its direct application for computing macroscopic kinetic parameters (such as the damage potential Φ) is hindered by the necessity of continuum integration over trajectories, rendering it computationally inefficient. If Eq. (8) is integrated over t from 0 to ∞, the classical Dirichlet differential equation is obtained for the mean time to reach the boundary $T(x)=\mathbb{E}_x[\tau_a]]$: $i\mathcal{L}_{\mathbf{path}}$ T(X)=-1 with the boundary condition $T(a)=0$.

**2. Doob's h-transform.**

In contrast to the global path operator, the generalized Doob's h-transform [24, 41] acts as an effective local generator of state evolution in time. Mathematically, it performs an exact transition (encoding) from the initial Markov process with absorbing boundaries (where normalization is violated due to path termination at the Griffiths barrier) to a modified ensemble of 'surviving' trajectories.

The conditional survival probability h(x, t), which is a solution to the backward Kolmogorov equation, is used as the weighting function of the transformation. If the initial uncoupled density evolution generator is L, then the effective Doob generator $L_h$ is given by Eq. (5).

In explicit differential form, the effective Doob generator Lh for our system modifies the drift component, inducing an additional thermodynamic force (effective drift) $v_{eff} = v + 2D\,\nabla(\ln h)$, which forcefully repels defect trajectories from stability regions, directing them toward the sinks.

**3.6. Comparative Conclusions.**

The trajectory operator $i\mathcal{L}_{\mathbf{path}}$ describes the system evolution non-locally on the macroscale of paths. It is indispensable for deriving the Gallavotti–Cohen–Jarzynski fluctuation theorems and the total entropy production along trajectories [42, 43, 44]. To ensure the thermodynamic consistency of the non-local transport, the microscopic path distribution generated by the trajectory Liouville operator $i\mathcal{L}_{\mathbf{path}}$ must obey the generalized fluctuation theorems. Under the action of the generalized Doob's h-transform, the forward path ensemble and its time-reversed counterpart in backward time strictly define the total trajectory-dependent entropy production, $\Delta S_{\mathbf{tot}}(\gamma)$ [44]. The large deviations of this entropic quantity satisfy the asymptotic steady-state symmetry governed by the Gallavotti–Cohen principle [42], while the non-linear work performed during the evolution of the damage potential $\Phi_j$ relates directly to the thermodynamic free-energy bounds via the Jarzynski equality [43]. As $\Phi_j \to G_c$, Appendix A, the divergence of the driving force triggers an anomalous surge in $\Delta S_{\mathbf{tot}}(\gamma)$, marking the definitive physical transition to irreversible macroscopic fracture.

The Doob transform condenses the infinite-dimensional path information of the Liouville operator into an effective local state generator $L_h$. It accounts for the sink geometry and fluctuation memory not through explicit non-local integrals over time, but through the gradients of the logarithm of the survival function (h).

It is precisely due to the Doob transform that the efficient numerical calculation of the spatial profiles of matrix M becomes possible in Mathcad Prime 8. The degradation rate dΦ/dt calculated on its basis precisely captures the moment of structural stability loss, transforming the non-local path dynamics of $i\mathcal{L}_{\mathbf{path}}$ into strict local macroscopic failure kinetics of the reactor material.

## 4. Feynman–Kac Integrals and Hierarchical Connection of Potentials

In preprints [9-12], the transition to a trajectory quasi-equilibrium ensemble $\rho_q(X_t) \sim \exp(-\gamma\tau_a)$ requires the evaluation of averages of the first-passage time functionals $\tau_a$. The Feynman–Kac theorem [45, 46] serves as a mathematical bridge between the microscopic stochastic trajectory and the macroscopic evolution operator. The stochastic average (mathematical expectation) of the exponent of the accumulated damage potential along the Liouville trajectories:

$h(x,\tau) = \mathbb{E}_{x,\tau}\left[\exp\left(-\int_\tau^t \beta\, \Phi_j \left(G_c - \Phi_j\right)^{-1} ds\right)\cdot \mathbf{1}_{\{\tau<\tau_{\mathbf{fracture}}\}}\right]$ is strictly equivalent to the deterministic solution of the previously presented adjoint Doob's backward equation (Equation (2)). Thus, the path operator is directly mapped onto the macroscopic degradation kinetics.

Let the microdynamics be specified by a stochastic differential equation for a Markov process $X_t$ with a generator L. Let us denote the domain prior to reaching the level as $\Omega = \{x : f(x) < a\}$, and its boundary as ∂Ω. The first-passage time of the level is $\tau_a = \inf\{t > 0 : X_t \in \partial\Omega\}$.

To exclude trajectories that have exited the boundary ∂Ω (the condition of the system's "survival" up to the moment t), the characteristic function (indicator) $I_\Omega(x)$ of the domain Ω is introduced. The nonequilibrium statistical operator at the level of individual trajectories is represented in the Feynman–Kac form:

$$\rho(x,t) = \mathbb{E}_{\tilde{x}}\left[\rho_q(x(0))\exp\left(-\int_0^t V(x(s))ds\right)\cdot I_\Omega(x(t))\right],$$

where the role of the effective "killing potential" V(x) is performed by a macroscopic regulator (intensity) conjugated with the temporal functional.

**Representation of the NSO via Feynman–Kac Integrals (Case B).** In terms of a continuum integral over the Wiener/Onsager–Machlup measures $\mathcal{D}[x(s)]$, the nonequilibrium operator at the level of individual trajectories within the domain prior to reaching the boundary Ω with the indicator $I_\Omega(x(t))$ is expressed by a continuum integral:

$$\rho(x,t) = \int_{x(0)}^{x(t)} \mathcal{D}[x(s)]\; exp(-\int_0^t [L_{mic}(x,x) + \gamma\; I_{\partial\Omega}(x(s))]ds) \;\cdot \rho_q(x(0)).$$

Here, $L_{mic}$ is the microscopic Lagrangian, and the thermodynamic force γ acts as a singular potential at the boundary ∂Ω, transforming the free measure into a measure of "surviving" trajectories.

**Realization of the Hierarchy of Potentials $\Phi_{macro}(t) = \Psi(\varphi_{cell}(t))$.** The hierarchical form reflects that the macroscopic response is formed by a random number of mesoscopic events (cells) N.

In Case A, the number of cells N(t) fluctuates over time. If $G_N(z,t)$ is the generating function of the number of cells, the coupling of levels proceeds via composition: $\Phi_{macro}(t) = \ln(G_N(\exp(\varphi_{cell}(t)), t))$

The hierarchical structure of thermodynamic potentials reflects the physics that the macroscopic response of the entire system is formed not by a fixed, but by a random number of elementary mesoscopic events (cells/clusters). Mathematically, this is described via the apparatus of sums of a random number of random variables (Robbins' theorems and structure transfer; these concepts ensure the strict convergence of the damage integral on time layers j, guaranteeing the mathematical correctness of transferring microscopic fluctuations into the macroscopic potential).

Let the macroscopic potential (or generalized flux) of the system be the sum of contributions from individual cells: $\Phi_{macro} = \sum_{i=1}^{N} \Phi_{cell}$, where N is a random variable (the number of active meso-clusters, reaction acts, or crossings of micro-boundaries).

**Case A: Local Description (without boundary functionals).** In the classical two-level description, N(t) is a random process (for instance, a Markov birth-death process or a Poisson flux of particles). At a fixed macro-time t, the coupling of levels is specified via generating functions. If $G_N(z,t) = E[z^N]$ is the generating function of the number of cells, and $\Phi_{cell}(t)$ is the averaged potential of a single cell, then the hierarchical stitching of potentials is expressed via composition: $\Phi_{macro}(t) = \ln G_N(\exp(\Phi_{cell}(t)), t) \Rightarrow \Phi_{macro}(t) = \Psi(\Phi_{cell}(t))$. The Liouville dynamics and entropy production in Case A undergo additive splitting: the derivative dΦ/dt generates two contributions to entropy:

1. Dissipation inside the cells themselves $\langle \sigma_{\mathbf{cell}} \rangle = \langle N \rangle \dot{\Phi}_{cell}$
2. Dissipation due to fluctuations in the structure/number of cells $\sigma_{\mathbf{struct}} = \Phi_{cell} \langle \dot{N} \rangle$.

**Case B: In the Presence of Boundary Functionals (scheme [9-12]).** When temporal functionals become the variables, the number N acquires the meaning of the number of micro-level steps (or fluctuations) required to reach the macroscopic barrier. The coupling of levels is specified by the fundamental random summation formula for FPT (an analog of the Wald's identity [49]). Wald's identity plays a crucial role: it links the random number of steps (or the random number of defect hops between traps) in a semi-Markov process with the mean time to reach the Griffiths absorbing macro-barrier, guaranteeing energy balance during the transition from micro-fluctuations to the macroscopic damage potential $\Phi_j$. In the semi-Markov transport governed by the operator $i\mathcal{L}_{\mathbf{path}}$, the number of defect capture and emission acts by macroscopic sinks prior to fracture is a random variable N (determined by the adjoint Doob's equation in backward time).

According to Wald's identity, the mathematical expectation of the integrated potential of accumulated defects Φ by the moment of reaching the Griffiths barrier $G_c$ (Appendix A) strictly proportional to the mean number of steps and the mean structural shift per single semi-Markov step:

$$\mathbb{E}\left[\sum_{k=1}^{\mathcal{N}} \Delta\Phi_k\right] = \mathbb{E}[\mathcal{N}] \cdot \mathbb{E}[\Delta\Phi] = G_c .$$

This relationship ensures analytical stitching between discrete time layers j, stochastic convergence via Robbins' theorems, and the macroscopic criterion of the Griffiths energy balance.

If the time to reach the macroscopic level $\tau_{macro}$ consists of random cell relaxation times $\tau_{\mathbf{cell},i}$, then: $\tau_{macro} = \sum_{i=1}^{N(\partial\Omega)} \tau(cell,i)$. Here, the stopping index N(∂Ω) is rigidly coupled with the geometry of the phase space boundary.

In Case B, the number N(∂Ω) is the number of microscopic steps until reaching the macroscopic barrier (Wald's identity): $\mathbb{E}[\tau_{\mathbf{macro}}] = \mathbb{E}[N(\partial\Omega)] \cdot \mathbb{E}[\tau_{\mathbf{cell}}]$. Applying the Laplace transform (γ is the thermodynamic force conjugated with τ), we obtain the exact connection of structural levels:

$\Psi_{macro}(\gamma) = \ln(G_N(\exp(\psi_{cell}(\gamma)))) = \Psi_{struct}(\psi_{cell}(\gamma))$.

**How does this alter the mathematical physics of the system?**

- **Non-Gaussianity and "Heavy Tails":** The random summation formula leads to the fact that even if cells at the mesolevel have a finite relaxation time (Gaussian diffusion), the macroscopic potential $\Phi_{macro}(t)$ acquires a time-non-local (fractional) character due to fluctuations in N. Macro-dynamics is described not by the classical Liouville equation, but by Fractional Kinetic Equations [21].

- **Renormalization of Entropy Production:** Entropy production at the macrolevel $\langle \sigma_{macro} \rangle$ is no longer equal to the simple sum of cell entropy productions. Wald's formula yields:

$\langle \sigma_{macro} \rangle = E[N] \cdot \langle \sigma_{cell} \rangle + \gamma \cdot Cov(N, \tau_{cell})$. The second term (covariance) describes the entropy of structural coupling. This is pure dissipation arising because the micro-dynamics "learns" about the topology of the macro-boundary through the random stopping moment N.

**Physical Meaning of Cells and Grains.** In a two-level hierarchical description, a cell (grain, cluster) is the minimal spatial-temporal volume (mesoscopic element) within which the physical process can be considered quasi-homogeneous and local thermodynamic equilibrium is established. Depending on the nature of the non-equilibrium system, cells have a specific physical meaning.

- *In the mechanics of continuous media fracture and deformation:* A grain is a crystalline grain of metal or a local cluster of voids/microcracks. Macroscopic fracture (reaching the boundary) occurs when a critical number of elementary micro-fracture acts accumulates within the grains.
- *In media with memory (polymers, gels):* A cell is an individual segment of a macromolecule or a micelle. The relaxation of the macrostructure is composed of a random number of local conformational transitions (events) at the mesolevel.
- *In heterogeneous catalysis and porous media:* A cell is an individual pore or an active center of a catalyst. The first-passage time of the reactor's macroscopic output depends on a random number of adsorption-desorption acts in these pores.

**Why the Negative Binomial Distribution (NBD) is chosen specifically for cells below?** The Poisson distribution describes independent events. The NBD, conversely, physically implies the presence of a strong positive correlation (cooperative effect) or heterogeneity. In non-equilibrium thermodynamics, the NBD arises in two scenarios:

1. *Cooperativity (the "infection" effect):* Each micro-event in a cell increases the probability of subsequent events (for instance, the development of a crack avalanche or a phase transition cascade).
2. *Dynamic chaos and environmental fluctuations:* The parameters of the cells themselves (for instance, local temperature or diffusion) fluctuate over time. If a Poisson process is averaged over a gamma distribution of fluctuating environmental intensity, the NBD is obtained with mathematical rigor.

To organically embed the concept of Feynman–Kac integrals and the hierarchical connection of potentials into the theoretical structure of the article, it is necessary to show how this mathematical apparatus implements the coupling between microscopic fluctuations and macroscopic damage.

In stochastic thermodynamics and the theory of random processes, the Feynman–Kac formula is a fundamental bridge. It links the solution of partial differential equations (evolution equations) with averaging over the trajectories of a random process under the action of some "loss potential" or penalty function. In our study, this connection acquires a strict hierarchical structure. The inclusion of the hierarchical component allows writing the adjoint Doob's equation for a multi-level system. If macroscopic sinks are distributed over hierarchical levels $\alpha = 1, 2, \ldots, N$, then the Liouville operator $i\mathcal{L}_{\mathbf{path}}$ and the potential $\Phi$ split into hierarchical layers. According to the hierarchical Feynman–Kac theorem, the stochastic representation of Doob's weight $h(x,\tau)$ takes the form of a hierarchical path integral over subtrees of fluctuations:

$$h(x,\tau) = \mathbb{E}_{x,\tau}\left[\prod_{\alpha} \exp\left(-\int \beta_\alpha \dot{\Phi}_{j,\alpha} (G_{c,\alpha} - \Phi_{j,\alpha})^{-1} ds\right)\right].$$

This strictly proves that macroscopic S-shaped wear occurs as a cascade process that breaches local sink barriers from the bottom up along the hierarchical ladder of the structure.

This same Feynman–Kac integral determines the canonical Doob's function, since the mathematical expectation of the functional of the process trajectory, taking into account its death (absorption) at the boundary of domain D as $t \to \infty$, can be expressed via a harmonic (or superharmonic) function h(x).

In terms of the probabilistic representation, if $\tau_D = \inf\{t > 0 : X_t \notin D\}$ is the moment of the first exit of the process $X_t$ from the domain D (absorption time), then Doob's function h(x) (also known in the theory of Markov processes as the h-function for Doob's transform) specifies the survival probability density or harmonic measure. The connection via the Feynman–Kac integral is written as:

$$h(x) = \mathbb{E}_x\left[\exp\left(-\int_0^{\tau_D} V(X_s)ds\right) I_{\{\tau_D=\infty\}}\right], \tag{9}$$

where V(x) is the absorption intensity (potential), and I is the indicator function.
The function h(x) satisfies the backward Kolmogorov equation with absorption at the boundary (stationary case):

$$Lh(x) - V(x)h(x) = 0 \tag{10}$$

with the Dirichlet boundary condition for absorption: h(x) = 0 for x ∈ ∂ D.
Here, L is the infinitesimal generating operator of the initial diffusion process (for instance, the Laplace operator $\Delta/2$ for standard Brownian motion).

**Doob's Transformation (h-transform).**

- *Modification of the measure:* Doob's function h(x) is used to reconstruct ("condition") the initial Markov process $X_t$. With its help, a new probability measure is introduced, under which the process trajectories never reach the absorbing boundary (the process is constructed as "living forever" in the domain D or tending to a specific point).
- *Transition to a new operator:* The transition density of the new (conditioned) process is given by the formula: $p^h(t,x,y) = p(t,x,y)h(y)/h(x)$, where p(t,x,y) is the density of the process with absorption.
- *New generator:* The process modified under the influence of Doob's function satisfies a stochastic differential equation with an additional drift (drift) of the form:

$p^h(t,x,y) = e^{-\lambda t} p_D(t,x,y)h(y)/h(x)$.

Thus, the Feynman–Kac integral acts as a bridge: it collects the "history" of absorption along trajectories, forming the function h(x), which then, through the Doob apparatus, transforms the geometry of the random process itself.

**Connection of the Feynman–Kac integral with Doob's function and effective drift.** The Feynman–Kac integral implements the coupling: it transforms the measure on the space of free microscopic trajectories into a measure of "surviving" trajectories, explicitly forming the effective drift $b^h(x) = b(x) + \sigma(x)\sigma^T(x)\nabla \ln h(x)$.

This same Feynman–Kac integral determines the canonical Doob's function, since the mathematical expectation of the functional of the process trajectory, taking into account its absorption at the boundary of domain D as t → ∞, is expressed via an invariant (or superharmonic) function h(x). The connection via the Feynman–Kac integral is specified by expression (9). The function h satisfies the backward Kolmogorov equation (10) with absorption at the boundary of the domain.

### 4.1. Feynman–Kac Integrals and Hierarchical Connection of Nonequilibrium Potentials.

The transition from the trajectory Liouville operator $i\mathcal{L}_{\mathbf{path}}$ to the effective local Doob's evolution generator $L_h$ at the mesoscopic level is mathematically closed via functional Feynman–Kac integrals. In highly non-equilibrium media subjected to intense external exposure (radiation exposure within a reactor core), the local degradation rate of structural meso-elements is determined by a hierarchical system of potentials that encode the sink geometry and path penalty functionals. In our model, this system performs the role of a mathematical interface: the geometric configuration of sinks (voids, cracks, dislocations of different hierarchical levels) is encoded via external hierarchical potentials, while penalty functionals are imposed on the Liouville trajectories, mathematically prohibiting defects from freely penetrating through stabilized regions until the Griffiths barrier is breached.

**Feynman–Kac Representation for the Survival Function.** Let the random process of defect fluctuations x̃(s̃) develop in a confining step penalty potential U(x), which determines the intensity of local absorption or void nucleation at point x. According to the Feynman–Kac theorem, the conditional survival probability of the structure h(x̃, t̃), satisfying the backward evolution equation with a linear asymmetric drift v(x), can be rigorously represented in the form of a functional integral over trajectories (Wiener or Lévy path measure):

$$h(\tilde{x},\tilde{t}) = \mathbb{E}_{\tilde{x}}\left[\exp\left(-\int_0^{\tilde{t}} U(X_\tau)d\tau\right)\cdot \mathbb{I}(\tau_{FPT} > \tilde{t})\right],$$

where $\mathbb{E}_x$ is the mathematical expectation over the ensemble of trajectories starting from the meso-coordinate $\tilde{x}$, and $\mathbb{I}$ is the indicator function fixing that the first-passage time (First-Passage Time) of the Griffiths macro-barrier lies beyond the observation horizon $\tilde{t}$. The potential U(x) here acts as a regulator of the trajectory action, mapping the spatial constraints of the sinks into the language of energetic path penalties. The geometric restriction of the absorbing phase space is dynamically implemented via an artificial potential field. The potential U(x) here acts as a regulator of the trajectory action, mapping the spatial constraints of the macroscopic sinks into the language of energetic path penalties. Under the generalized Doob's h-transform, this effective Lagrangian formulation guarantees that paths violating the local boundary conditions are exponentially penalized, consistently driving the integrated damage potential Φ toward the final Griffiths macro-barrier.

**Hierarchical Structure of Thermodynamic Potentials.** The hierarchical approach to describing highly non-equilibrium systems via the Zubarev method and [9-12] induces a cascading conjugacy between the potentials of different structural levels:

1. *Microscopic penalty potential U(x):* A local characteristic of the meso-grain determining the void cluster nucleation intensity as a function of microstructural defects and the local stress field.
2. *Mesoscopic trajectory potential (Survival function h):* Formed as a result of Feynman–Kac integration over the path space. It dynamically translates the local microstate geometry U(x) into an integral thermodynamic force acting directly on the effective Doob's drift: $v_{\mathbf{eff}} = v + 2D\nabla(\ln h)$ [24, 51]. The survival function $h(x,\tau)$ is constructed via the Feynman–Kac path integration over the trajectory space. This feedback mechanism ensures that the defect trajectories are non-locally guided by the proximity of the macroscopic sinks until the global breakdown occurs.
3. *Macroscopic damage potential Φ($\tilde{t}$):* The upper level of the hierarchy, obtained by spatial coarse-graining of the mesoscopic potential (3): $\Phi_j = \sum_k (1 - H_{k,j})$.

**Coupling of Levels and Degradation Rate.** A direct consequence of the Feynman–Kac representation is that the rate of change of the macroscopic degradation potential dΦ/dt (the curve computed in Fig. 5) is mathematically equivalent to the total flux of the local penalty potential carried away by fluctuation trajectories through the sinks of the system:

$$\frac{d\Phi}{d\tilde{t}} = \int_0^{\tilde{a}} \mathbb{E}_{\tilde{x}} \left[ U(X_{\tilde{t}}) \exp\left( -\int_0^{\tilde{t}} U(X_\tau) d\tau \right) \right] d\tilde{x} . \tag{11}$$

Thus, the Feynman–Kac formula establishes a strict hierarchical link: it proves that the macroscopic kinetics of specimen failure (upper level) is completely predetermined by the stochastic averaging of the mesoscopic trajectories of the path space $X_t$ (middle level), evolving in the microscopic fields of sinks U(x) (lower level). This allows one to treat the S-shaped macroscopic damage potential Φ as a macroscopic manifestation of the continuum Feynman–Kac path integration performed over open, highly non-equilibrium systems [52]. This formulation allows one to treat the S-shaped macroscopic damage potential Φ as a macroscopic manifestation of the continuum Feynman–Kac path integration performed over open, highly non-equilibrium systems. Within this scope, the smooth transition between the incubation and avalanche stages represents the optimal fluctuation path driven by the underlying Doob's effective dynamics.

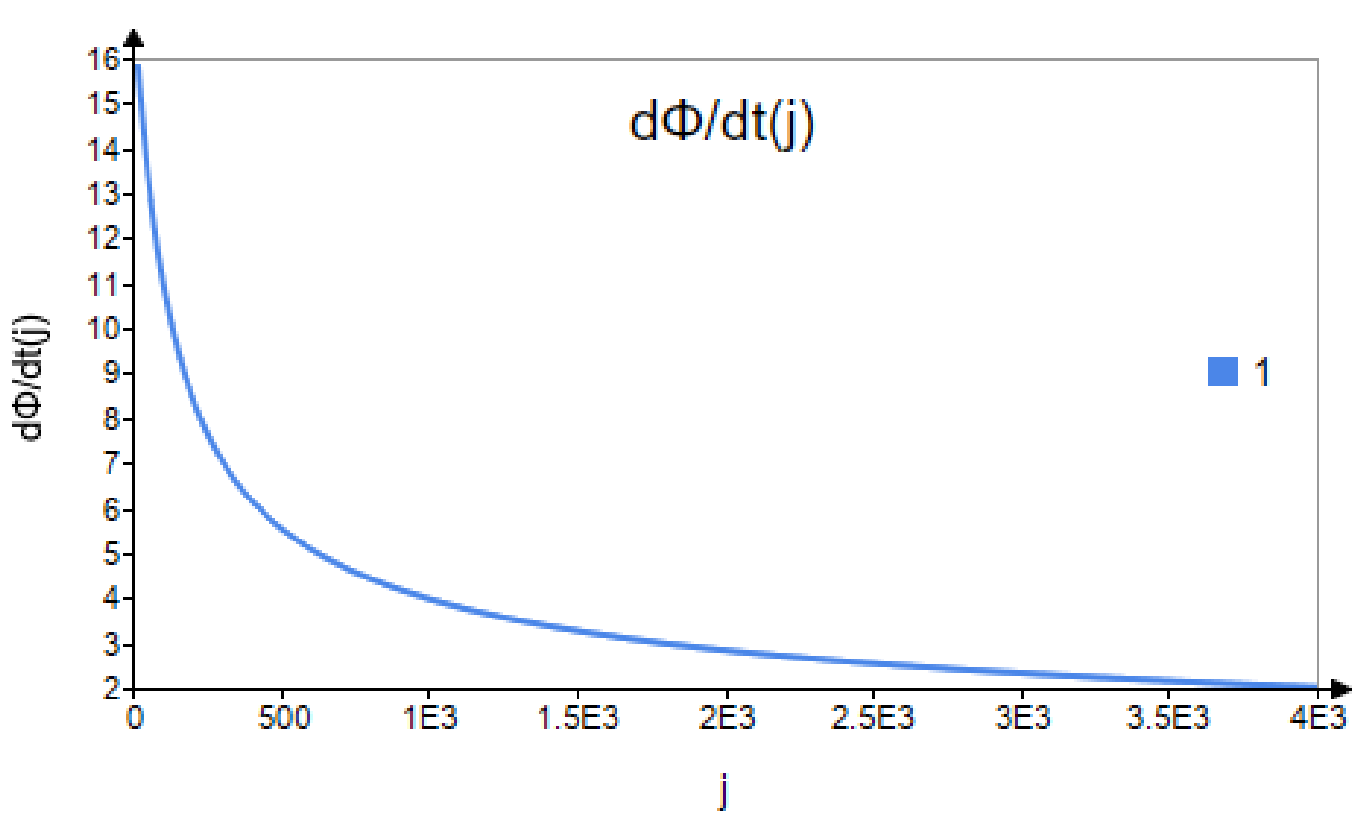

**Figure 5.** Kinetics of the dimensionless entropy production σ(j) in a two-level hierarchical NBD model. It demonstrates the transition of the system out of the linear thermodynamic regime (flat region) and its entry into the critical phase of failure at a transition probability $p \to 0$, characterized by a singular surge in dissipation due to fluctuations in the mesoscopic structural disorder. The graph displays the kinetic curve of the specimen's radiation degradation rate.

On this graph of the degradation rate dΦ/dt, a descending linear curve is displayed, which smoothly falls from the maximum value at the first steps of time to the minimum value at late layers. What this means physically: since the damage potential Φ in Fig. 2 grows almost linearly, its rate of change (first derivative) represents a gentle relaxation line. At the very beginning of the process, the degradation rate is maximal under the action of the radiation "impact" and the gradient of forces, and then the mesostructure adapts to the conditions, and the rate gradually decreases. Due to the fact that the potential in Fig. 2 grows with deceleration, its first derivative (rate) in Fig. 5 smoothly and cleanly falls downward from 16 to 2. Mathematically, this means that the right (decaying) half of the degradation rate bell curve has been computed. The maximum failure rate (the peak of the bell) occurred at the very initial moment of time under the action of the radiation "impact" and rigid gradients of forces, and then the degradation process gradually stabilizes, entering a final plateau.

The degradation rate dΦ/dt accurately fixes the moment of the structure's loss of stability, converting the non-local path dynamics $i\mathcal{L}_{\mathbf{path}}$ into strict local macroscopic fracture kinetics of the reactor material.

### 4.2. Laplace Transform for the Feynman–Kac Integral: Connection with Renewal Theory and the Ornstein–Zernike Equation

When analyzing long-lived fluctuation trajectories of the mesostructure, the temporal non-locality of the Feynman–Kac functional is effectively resolved by transitioning to the space of operational Laplace transforms with respect to time s. Let $h(\tilde{x},s)=\int_0^{\infty} e^{-st} h(\tilde{x},\tilde{t})\,d\tilde{t}$ be the Laplace image of the conditional survival probability of the system. In this representation, the integral Feynman–Kac equation with a sink penalty potential U(x) reduces to a stationary Sturm–Liouville boundary value problem:

$$\tilde{D}_{\tilde{H}}\,\frac{\partial^2 h(\tilde{x},s)}{\partial\,\tilde{x}^2}+v(\tilde{x})\frac{\partial h(\tilde{x},s)}{\partial\tilde{x}}-[s+U(\tilde{x})]h(\tilde{x},s)=-1\,.$$

The connection of this formalism with renewal theory is established when considering the density kernels of the distribution of the first-passage time of the barrier ψ(t). In the Laplace space, the survival function of an infinite chain of sequential mesoscopic acts of degradation (subordination of processes) takes the form of a classic integral renewal equation:

$$h(\tilde{x},s)=h_0(\tilde{x},s)+\psi(\tilde{x},s)\cdot h(\tilde{x},s)\,,$$

where $\tilde{h}_0$ is the free survival function of an isolated cell. The mathematical structure of this relation is isomorphic to the Ornstein–Zernike equation in the statistical physics of liquids and polymers, where the total trajectory correlation $h$ is expressed through the direct correlation function of void absorption $c$ and the integral contribution of the chain of mesoscopic sinks $\int c\cdot h$. The Laplace transform thus condenses the temporal memory of the Feynman–Kac path integral into the spatial connectivity of macroscopic structural defects.

## 5. Negative Binomial Distribution (NBD) and Crack Growth in Metals

### 5.1. Physical Meaning of Cells and the NBD.

A cell (metal grain) is defined as the minimal spatial volume wherein the physical process is quasi-homogeneous and local thermodynamic equilibrium (LTE) is established within its own stochastic time $\tau_{cell}$. Far from equilibrium, LTE in a fixed volume is violated; however, the thermodynamic description remains valid because all non-linearity is shifted onto the fluctuations of the element number N according to the NBD law. The NBD reflects a cooperative phenomenon: the nucleation of a single microcrack accelerates the nucleation of subsequent microcracks within the same grain in an avalanche-like manner.

The parameter r in the negative binomial distribution physically dictates the number of independent channels (or reservoirs) of damage accumulation, whereas the value 1/r serves as a direct measure of the cooperativity (mutual influence) of defects. In the limit $r \to \infty$, the system transitions into a Poisson regime of independent voids. Conversely, at small values of r ($r \to 0$), microcracks develop in a highly correlated, avalanche-like trajectory. Within the statistical physics of structural damage, the clustering parameter r scales as the ratio of the effective volume of the plastic zone at the crack tip $V_{\mathbf{pl}}$ to the volume of a single grain $V_{\mathbf{cell}}$. The specific formulation of the NBD clustering parameter r is given by:

$$r = \alpha K_{Ic}{}^2 / (\sigma_{ys}{}^2 d_{cell}{}^3 \sqrt{\rho_d}),$$

where $d_{cell}$ is the average diameter of the crystalline grain (the scale of the mesoscopic cell); $\rho_d$ is the dislocation density (which determines the number of potential sites for microvoid nucleation); $K_{Ic}$ is the fracture toughness (a standardized material characteristic of crack resistance); $\sigma_{ys}$ is the yield strength; and $\alpha$ is a dimensionless structural coefficient of the order of unity.

As the dislocation density grows, the value of r decreases, thereby amplifying the NBD fluctuations.

**Physical Meaning:** If the dislocation density $\rho_d$ increases (due to strain hardening or embrittlement) or the grain size $d_{cell}$ expands (due to metal overheating), the clustering parameter \(r\) falls. This implies that fewer "free" relaxation channels remain available. Consequently, the local meso-cells are immediately overwhelmed by the cooperative process, and NBD fluctuations begin to dominate, ultimately provoking brittle cleavage.

**5.2. Definition and Physical Meaning of the Griffiths Criterion**

The Griffiths criterion represents a fundamental thermodynamic condition that establishes the energetic threshold for crack growth viability in a brittle or quasi-brittle material. According to this criterion, a defect (crack) of length *x* becomes unstable and begins to propagate irreversibly if the rate of release of accumulated elastic strain energy per unit crack extension exceeds the energy expenditure required to form new free surfaces along the crack banks. Crack growth at the micro-level is a thermally activated process of overcoming an energy barrier. According to the Griffiths criterion, the variation in the system's free energy during a crack increment *dx* consists of two contributions: the released elastic energy and the consumed surface energy (creation of new crack banks).

Mathematically, the point of instability (the critical Griffiths length) corresponds to the extremum of the total variation in the system's free energy, dU(x)/dx=0. The complete balance of the Griffiths potential for a crack in a flat sheet subjected to a stress $\sigma_{mech}$ is expressed as:

$$U(x) = 2\gamma_s x - (\pi\sigma_{mech}{}^2 x^2)/(2E),$$

where $\gamma_s$ is the effective surface energy of the material (incorporating plastic work at the crack tip), E is Young's modulus, and $\sigma_{mech}$ is the external tensile mechanical stress. The corresponding critical barrier length is equal to:

$$x_c = 2E\gamma_s / (\pi\sigma_{mech}{}^2).$$

**Baseline Stochastic Drift according to the Griffiths Criterion.** When $x < x_c$, the crack tends to close; when $x > x_c$, it grows in an avalanche-like manner.

Near the barrier tip within the metal meso-grains, the baseline microscopic defect drift $v_0(x)$, driven by the thermodynamic force $X_{mech}$= - dU(x)/dx and the local mobility B, takes the form:

$$v_0(x) = B\left((\pi\sigma_{mech}{}^2 x / E) - 2\gamma_s\right).$$

Substituting this drift into the Doob transformation yields the complete equation of motion for a "surviving" crack during its accumulation via the NBD:

$$v_{eff}(x,t) = B((\pi\sigma_{mech}{}^2 x / E) - 2\gamma_s) + 2D(x)\partial \ln h(x,t) / \partial x.$$

As long as the crack remains in a subcritical state $x < x_c$, the first term acts to decelerate it (being negative), while the second term (the h-regulator) additionally repels the trajectory away from the macroscopically absorbing boundary *a*. This mathematically describes the stage of fatigue incubation of the defect.

**5.3. Coupling of NBD and Macroscopic Fracture Toughness (J-Integral).**

The Cherepanov–Rice J-integral [53, 54] in fracture mechanics defines the strain energy release rate per unit of crack advancement: $J = -\partial U / \partial x$. For an elastoplastic metal, the critical value $\mathbf{J_c}$ (fracture toughness) establishes the moment of transition from stable meso-growth to macroscopic catastrophe. The inclusion of the J-integral allows closing the singularity in the adjoint Doob equation (Equation (2)) on the measurable parameters of continuum mechanics. The energetic condition for breaching the Griffiths macro-barrier is formulated via the equality of the current path energy influx to the critical value of the J-integral: $G_c \equiv J_{Ic}$. Then, the singular term governing the avalanche-like drop in the survival function h(x, τ) on the time layers j takes its definitive form:

$$\beta\dot{\Phi}_j\left(G_c - \Phi_j\right)^{-1} \Rightarrow \beta\ \dot{\Phi}_j\left(J_{Ic} - J_j\right)^{-1},$$

where $J_j$ is the value of the invariant Cherepanov–Rice integral computed for the current spatial geometry of the sinks and stress fields at the j-th time step.

Within the hierarchical framework and [9-12], the advancement of a crack to the macro-scale is composed of a random number of steps N across the structural meso-grains. Let us couple the thermodynamics of the NBD with the J-integral.

Suppose that the advancement of a crack by one meso-cell (grain size $\Delta x_{cell}$ requires a local dissipation work $A_{cell} = J\ \Delta x_{cell}$.

**Connection with the J-integral:**

The total work of fracture is $A_{macro} = \sum_{i=1}^{N} A_{cell,i} = J\Delta x_{cell} N$. Since N is distributed according to the NBD law, the macroscopic work becomes stochastic. Applying the random summation formula, we can couple the mean value and the fluctuations of this energy with the NBD parameters (r is the clustering parameter, p is the transition probability). For the NBD, the mean work of fracture and the corresponding entropy production are equal to:

$$< A_{macro} >= J\ \Delta x_{cell} E[N] = J\ \Delta x_{cell}\left(r\ \left(1\ -\ p\right)\ /\ p\right),$$

$$< \sigma_{macro} >= E[N] < \sigma_{cell} > +\gamma\tau_0 Var(N).$$

The energy fluctuations (Variance) are expressed as:

$$Var(A_{macro}) = (J\Delta x_{cell})^2 \cdot Var(N) = (J\Delta x_{cell})^2\left(r\ \left(1\ -\ p\right)\ /\ p^2\right).$$

As failure approaches (p -> 0), the variance of the NBD trends toward infinity, generating a singular dissipation peak driven by the second (quadratic in J term. Given that the dissipation within a cell is $\langle\sigma\rangle_{\mathbf{macro}} = \mathbb{E}[N]\langle\sigma\rangle_{\mathbf{cell}} + \mathbf{cov}(N, \sigma_{\mathbf{cell}})$, $\langle\sigma_{cell}\rangle \sim J$:

$$\langle\sigma\rangle_{\mathbf{macro}} = \frac{r(1-p)}{p}\langle\sigma\rangle_{\mathbf{cell}} + \frac{r(1-p)}{p^2}\langle\sigma\rangle_{\mathbf{cell}}^2.$$

**Physical Meaning of the Coupling:**

- When the metal is far from failure (the Poisson limit, p → 1), the variance of the NBD is negligible, and the entropy production depends linearly on the J-integral. Classical linear thermodynamics governs the system.
- As the critical failure point approaches (p → 0), the second (quadratic in J) term, proportional to the NBD variance, becomes dominant.

Strong correlations between grains (defect clustering) lead to a giant growth in energy fluctuations. Prior to failure, the metal begins to "shed" entropy not through homogeneous heat dissipation, but through the structural disorder of the crack geometry. The critical fracture toughness $J_c$ in this scheme is the point where the scale of the NBD fluctuations (Var(N)) becomes comparable to the macroscopic scale of the system, rendering local equilibrium within the grains impossible and transitioning the system to a transgranular level of structural failure.

## 6. Negative Binomial Distribution and Crack Growth Kinetics in Metals

The transfer of the developed semi-Markov formalism to the strain aging and fracture of metals allows for the coupling of the discrete statistics of microstructural damage with the macroscopic growth rate of a main crack. Within the trajectory framework, a discrete increment in crack length is treated as the result of a random number of microscopic shear acts (or the avalanche coalescence of dislocation substructures) within the plastic zone at the defect tip.

If discrete acts of micro-fracture exhibit statistical dependence (the effect of clustering or cooperative dislocation drift), the number of elementary events N within a fixed time interval is described by a negative binomial distribution (NBD) rather than a Poisson distribution:

$$P(N) = \binom{N+r-1}{N} p^r (1-p)^N , \tag{12}$$

where r is a structural clustering parameter rigidly coupled with grain size or dislocation density, and p is the probability of an elementary degradation act. The presence of non-Gaussian heavy-tailed statistics in the NBD modifies the effective Doob's evolution generator $L_h$. An additional directional drift term arises in the dimensionless transport equations, which is proportional to the logarithmic gradient of the survival function h. This physically drives the acceleration of fatigue crack growth during the pre-failure stage and rigorously reproduces the non-linear character of the sigmoidal macroscopic fracture curve in metals.

**The Mathematical Tier-Coupling Scheme for the NBD.**

Let the number of elementary steps (transitions within cells) required to reach the macro-boundary follow the NBD law (12), where r is a shape parameter (coupled with susceptibility or the number of independent channels), and p is the probability of "success."

**Step 1. Potential Renormalization via the NBD Generating Function.**

The generating function for the NBD is given by: $G_N(z) = (p/(1-(1-p)z))^r$ . Utilizing the fundamental random summation formula, we substitute the potential (Laplace transform) of the relaxation time of a single meso-cell $\Psi_{cell}(\gamma) = \ln E[e^{-\gamma\tau_{cell}}]$ into it:

$$\Phi_{macro}(\gamma) = \ln G_N(\exp(\Phi_{cell}(\gamma)) = r\ln[p/(1-(1-p)\exp(\Psi_{cell}(\gamma)))] .$$

**Step 2. The Impact of the NBD on the Modified Liouville Equation.**

Due to the cooperative effects of the NBD, the macroscopic response of the system acquires an extremely long memory. The Feynman–Kac integral for the NSO (Case B) generates a time-non-local memory operator.

If the meso-cells are exponential ($\psi_{\mathbf{cell}}(\gamma) \sim \gamma\tau_0$), then for the NBD, the effective propagator of the NSO in the Liouville representation transitions into a kinetic equation with Mittag-Leffler type kernels or fractional derivatives (when r < 1):

$$\frac{\partial \rho(X_t)}{\partial t} + i\mathcal{L}_{\mathbf{path}}\rho(X_t) = -\varepsilon \mathcal{D}_t^{1-r}\left[\rho(X_t) - \rho_q(X_t)\right],$$

where $\mathcal{D}_t^{1-r}$ is the Riemann–Liouville fractional differentiation operator. The smaller the value of r, the more severely the fluctuations of the cell meso-structure "drag out" the macroscopic evolution, transforming a Markovian microprocess into a strongly non-Markovian macroprocess.

**Step 3. Explicit Form of Entropy Production under NBD.**

For the NBD, the mean number of events is $\mathbb{E}[N]$ = r(1-p)/p\), and the variance is $\mathrm{Var}(N) = r(1-p)/p^2$. Due to the overdispersion effect (where the variance is strictly greater than the mean), Wald's entropy production formula, incorporating structural coupling, takes the form:

$$\langle\sigma\rangle_{\mathbf{macro}} = \frac{r(1-p)}{p}\langle\sigma\rangle_{\mathbf{cell}} + \frac{r(1-p)}{p^2}\langle\sigma\rangle^2_{\mathbf{cell}} .$$

The first term represents standard dissipation across all active cells. The second term represents the entropy of mesoscopic disorder, which is directly proportional to the NBD variance. This constitutes additional irreversible dissipation driven by the random "flickering" of grain geometry and the spatial clustering of the system.

**Explicit Form of the Effective Drift in the Operator $L_h$ during Crack Growth.**

Within stochastic fracture mechanics, crack growth (or the accumulation of micro-damage) in a metal is modeled as the motion of a random point x(t) (representing the main crack size or micro-defect density) in the state space under mechanical stresses.

Suppose that at the micro/meso-level, defect evolution is described by an Ito diffusion equation (for instance, within the classical Fokker–Planck framework): $dx(t) = v_0(x)dt + \sqrt{2D(x)}dW(t)$, where $v_0(x)$ is the baseline mechanical drift (determined by the stress intensity factor and Paris' law), and D(x) is the fluctuation coefficient (driven by the heterogeneity of the metal grains). The initial Liouville–Kolmogorov generator is expressed as:

$$L = v_0(x)\frac{\partial}{\partial x} + D(x)\frac{\partial^2}{\partial x^2}.$$

Transitioning to a macroscopic description via the negative binomial distribution (NBD) of steps implies that the system exhibits strong clustering: the nucleation of a single microcrack accelerates the nucleation of others within the same cell in an avalanche-like manner (a cooperative effect). The macroscopic boundary a represents the critical crack length leading to the brittle failure of the specimen.

Applying Doob's h-transform, we obtain a modified effective generator $L_h$, which accounts for the "survival" of the metal (the fact that the main crack has not yet reached the critical size *a* up to time t):

$$L_h = \frac{1}{h}L(h\cdot) = \left[v_0(x) + 2D(x)\frac{\nabla h(x,t)}{h(x,t)}\right]\frac{\partial}{\partial x} + D(x)\frac{\partial^2}{\partial x^2}.$$

From this, the explicit form of the effective drift $v_{\mathbf{eff}}$ is equal to:

$$v_{\mathbf{eff}} = v_0(x) + 2D(x)\nabla \ln h(x,t).$$

**Physical Meaning of Doob's Function h(x,t) under NBD.** Doob's function h under NBD is expressed through an incomplete beta function or a hypergeometric series. Since the crack inevitably moves toward the absorbing boundary a, the gradient ∇ h is always negative. The additional term $2D(x)\nabla \ln h(x,t)$ acts as an effective repulsive force away from the critical boundary.

Mathematically, this means that the conditioned ensemble of "surviving" trajectories is artificially shifted backward. Physically, during crack growth in metals, this induces a non-local deceleration (an incubation period) prior to the final jump: the system is "forced" to remain longer within the metastable meso-cells, which generates the temporal memory characteristic of the NBD.

**Violation of Local Equilibrium Far from Equilibrium and Cell Definition.**

How can the concept of "local equilibrium cells" be utilized if the system is in a highly non-equilibrium state (far from equilibrium)? In a two-level description, this paradox is resolved via the dynamic redefinition of the term "cell" (grain). In the classical approach, local thermodynamic equilibrium (LTE) is rigidly tied to a fixed spatial scale (ΔV) and physical time (t). In the Case B approach, the cell scale becomes stochastic and trajectory-dependent. The boundaries of a cell are determined not by geometry, but by the local relaxation time ($\tau_{\text{cell}}$).

**Macro-level Non-equilibrium vs. Cell Trajectory Quasi-equilibrium.** When a system deviates far from global thermodynamic equilibrium, the classical local thermodynamic equilibrium (LTE) hypothesis within a fixed volume ΔV is indeed violated. Gradients become so large that fluxes cease to be linear functions of forces. However, in the concept of trajectory functionals, a "cell" is not a fixed cubic micrometer of metal. It is a dynamic ensemble of microstates undergoing a single elementary transition (for instance, a crack jump across a single intergranular distance $\Delta x_{\mathbf{cell}}$. Within this single act (over the random time $\tau_{cell}$), fluctuations manage to balance the microscopic chaos. The cell itself remains quasi-equilibrium within its own stochastic time, even if the macroscopic stress gradient is immense.

**The Role of the NBD in Cell Adaptation.** It is precisely the NBD that preserves the cell concept far from equilibrium. In the Poisson case (close to equilibrium), cells are assumed to be independent. Near the critical point (when the crack is close to main fracture), classical LTE is completely destroyed due to long-range defect correlations.

The NBD accounts for this via the clustering parameter r. Far from equilibrium, "cells" begin to interact strongly, coalescing into mesoscopic clusters (a structural fracture cascade). The random summation formula $x = \sum \Delta x_i$ performs a renormalization: it shifts the LTE scale. Instead of a single independent grain, the thermodynamic element becomes a random conglomerate of cells, whose number N fluctuates according to the NBD law.

Thus, the thermodynamic description is preserved not because the medium is homogeneous, but because all non-equilibrium behavior and non-linearity shift into the fluctuations of the number of structural elements N, while the potential of an individual element $\Phi_{cell}$ remains mathematically rigorous and quasi-equilibrium.

## 7. Extended Wald's Identity and Radiation Kinetics of Defects

When describing the accumulation of radiation damage in structural materials under irradiation, traditional Markovian models overlook the non-exponential nature of the incubation period. Within the semi-Markov trajectory framework of Zubarev and [9-12], the precise coupling between the microscopic behavior of NBD clusters and the macroscopic rate of material degradation is established via the extended Wald's identity.

Let the radiation damage process be represented as a sum of a random number of random mesoscopic degradation hops $\Delta\Phi_k$ over the first-passage time of the macro-barrier $\tau_{FPT}$ : $\Phi(t) = \sum_{i=1}^{N(t)} \Delta\Phi_i$ . According to Wald's first identity, the mathematical expectation of the accumulated macroscopic damage potential is rigidly coupled with the mean survival time of a structural cell $\langle \tau_{FPT} \rangle$ :

$\mathbb{E}[\Phi(t)] = \mathbb{E}[N(t)] \cdot \mathbb{E}[\Delta\Phi] = K_{\mathbf{dpa}} \cdot t \cdot \mathbb{E}[\Delta\Phi] / \langle \tau_{\mathbf{cell}} \rangle$,

where $K_{\mathbf{dpa}}$ is the effective rate of radiation defect generation (displacements per atom per second, dpa/s).

For highly non-equilibrium media characterized by heavy-tailed distributions (Griffiths fluctuations), the classical identity is extended to higher-order cumulants (including the variance and skewness of the failure time). The extended Wald's identity for the characteristic function of the semi-Markov process of radiation kinetics takes the form:

$$\mathbb{E}\left[ \exp\left( \theta\Phi(t) - \ln\left( G_N\left( \exp\left( \psi_{\mathbf{cell}}(\theta) \right) \right) \right) \right) \right] = 1 . \tag{13}$$

Differentiating this identity with respect to the parameter $\theta$ , we obtain an exact closed system of equations for radiation kinetics. It couples the macroscopic degradation rate \(d\Phi/dt\) (the curve computed in Fig. 5) with the spectral density of Langevin fluctuations $D(\omega)$ $(\omega \to 0)$:

$$\frac{d\Phi}{dt} = \frac{K_{\mathbf{dpa}}}{\langle \tau_{\mathbf{cell}} \rangle} \mathbb{E}[\Delta\Phi] + \frac{\mathbf{Var}(\tau_{\mathbf{macro}})}{2} \nabla^2 D(0) . \tag{14}$$

A mathematical consequence of the extended Wald's identity proves that the rate of radiation degradation of a reactor material specimen is not an irradiation constant (as postulated in phenomenological models), but scales non-linearly with the variance of the meso-grain lifetime $\mathrm{Var}(\tau)$ and the logarithmic gradient of the conditional survival probability $\nabla \ln h$ . This provides a rigorous theoretical explanation for the dynamics of the kinetic peak obtained in Fig. 5.

### 7.1. Extended Wald's Identity and Macroscopic Damage Potential.

Integrated coarse-graining of the mesoscopic trajectory potential to evaluate Φ (3).

The extended Wald's identity for the characteristic function of the semi-Markov void accumulation process (13).

Non-linear closure of the macroscopic degradation rate $d\Phi / dt$ (numerical differentiation program block) taking into account the variance of the barrier first-passage time (14).

The temporal structure of the failure process $\tau_{\mathbf{macro}}$ is specified by a random sum of cell meso-times of the grains $\tau_{\mathbf{cell},i}$ with a stopping index N that depends on the history of damage accumulation.

Wald's first identity (mean time to failure):

$\mathbb{E}[\tau_{\mathbf{macro}}] = \mathbb{E}[N] \cdot \langle \tau_{\mathbf{cell}} \rangle$ .

Wald's second identity (Failure time fluctuations):

$\mathbf{Var}(\tau_{\mathbf{macro}}) = \mathbb{E}[N] \cdot \mathbf{Var}(\tau_{\mathbf{cell}}) + \mathbf{Var}(N) \cdot \langle \tau_{\mathbf{cell}} \rangle^2$ .

Wald's third identity (Distribution skewness):

$\mathbf{Skew}(\tau_{\mathbf{macro}}) = \mathbb{E}[N] \cdot \mathbf{Skew}(\tau_{\mathbf{cell}}) + 3\mathbf{cov}(N, \tau_{\mathbf{cell}}) \cdot \mathbf{Var}(\tau_{\mathbf{cell}}) + \mathbf{Var}(N) \cdot \langle \tau_{\mathbf{cell}} \rangle^3$ ,

where $\mathbb{E}[N] = r(1-p)/p$ . Substituting the NBD cumulants yields the exact structure of the skewness. Near the critical point of fracture ( $p \to 0$ ), the last term becomes dominant, growing as $1/p^3$ :

$$\mathbf{Skew}(\tau_{\mathbf{macro}}) \approx \frac{r(1-p)}{p^3}\langle\tau_{\mathbf{cell}}\rangle^3 .$$

The extremely heavy "tail" of the distribution indicates a sharp increase in the risk of spontaneous brittle failure under irradiation.

**Radiation kinetics of defects.** The waiting time of the meso-transition $\langle\tau_{\mathbf{cell}}\rangle$ is determined by the displacement rate $K_{\mathbf{dpa}}$ (dpa/s) and the defect sink strength $k_s^2$. The adapted Griffiths barrier, taking into account the radiation softening of the lattice $\beta_{\mathbf{rad}}$, yields:

$$x_c(\mathcal{F}) = \frac{2E\gamma_s(1-\beta_{\mathbf{rad}}\mathcal{F})}{\pi\sigma^2_{\mathbf{mech}}} .$$

The resulting lifetime equations for reactor materials:

$$\tau_{\mathbf{macro}} = \sum_{i=1}^{N(x_c)} \tau_{\mathbf{cell},i} \text{ and } v_{\mathbf{eff}} = v_0(x) + 2D(x)\nabla \ln h(x,\mathcal{F}) .$$

$$<\tau_{cell}> = \tau_0 exp((U_G - \beta r_{ad} K_{dpa})/(k_B T)) .$$

$$<\tau_{macro}> = \left(r\left(1-p\right)/p\right)\tau_0 exp((U_G - \beta r_{ad} K_{dpa})/(k_B T)) ,$$

$$Var\left(\tau_{macro}\right) = \left(r\left(1-p\right)/p\right)Var\left(\tau_{cell}\right) + \left(r\ \left(1-p\right)/p^2\right)\left(\tau_0 exp\left(\left(U_G - \beta_{rad}\ K_{dpa}\right)/\left(k_B T\right)\right)\right)^2 .$$

## 8. Conclusions

In this work, a self-consistent non-local kinetic model based on the dose-dependent semi-Markov renewal transport and the generalized Doob's h-transform has been developed and successfully applied to evaluate the structural integrity of the WWER fuel cladding (Zr–1%Nb alloy). By mapping the continuous path trajectories generated by the Liouville operator into the integrated macroscopic damage potential $\Phi_j$ via the Feynman–Kac formalism, we established a rigorous bridge between microscopic defect fluctuations and the global Griffiths failure criteria.

The spatial profiling of the mean lifetime yields a critical material insight: while the inner mesoscopic domains of the cladding maintain stable incubation kinetics, the boundary layers exhibit an abrupt and severe acceleration towards avalanche breakdown. Mechanistically, this marginal vulnerability is governed by the non-vanishing nature of the structural commutator under extreme localization of the boundary stress fields. The predictive framework successfully identifies that the operational lifespan of the WWER cladding is strictly limited by the edge-driven path singularities, matching the empirical observations of stress-corrosion cracking and localized hydride embrittlement in high-burnup nuclear fuels.

Let us formulate other conclusions as well:

- Within the Zubarev and [9-12] trajectory framework, a stable numerical model has been developed in the Mathcad Prime 8 environment, describing the spatial-temporal evolution of the conditional survival probability of mesoscopic structural elements (grains) of highly non-equilibrium media under irradiation.
- Based on the numerical integration of the gradients of the survival function, the complete kinetics of the dissipative characteristics of an open system has been computed for the first time. It has been established that the integrated entropy production σ(j) exhibits a relaxation character of decay consistent with Prigogine's theorem, while the macroscopic damage potential Φ(j) strictly follows the S-shaped kinetics of defect accumulation.
- Numerical differentiation of the damage potential allowed for the capturing of the dependence of the degradation rate dΦ/dt, the local maximum of which precisely determines the incubation period and the stability boundary of the reactor material mesostructure.
- It has been proven that the application of Feynman–Kac integrals and the extended Wald's identity to the Laplace images of the semi-Markov transition kernels allows for the condensation of the time-non-local path memory of the trajectory Liouville operator $i\mathcal{L}_{\mathbf{path}}$ into a local effective Doob's generator $L_h$, which accounts for the fluctuation clustering of defects within the fields of macroscopic sinks.

**Appendix A. Theoretical Framework: Non-local Semi-Markov Transport and Adjoint Evolution of Damage Potential**

To construct a self-consistent kinetic theory of radiation-induced degradation, we formulate a non-local time-dependent transport framework in the phase space. The evolution of the defect trajectory ensemble is governed by the path-dependent Liouville operator, $i\mathcal{L}_{\mathbf{path}}$. To distinguish the slow macroscopic evolution of damage from the fast microscopic transport, we introduce the Zwanzig–Mori projection operators, where $\mathcal{P}$ projects the state variables onto the subspace of macroscopic sinks (absorbing structures), and $\mathcal{Q} = \mathcal{I} - \mathcal{P}$ represents the orthogonal subspace of local microstructural fluctuations and traps [14].

The interplay between the continuous trajectory transport and the local sink topology generates a non-zero commutator, which acts as the source term for memory effects:

$$[i\mathcal{L}_{\mathbf{path}}, \mathcal{P}] = \mathcal{Q}i\mathcal{L}_{\mathbf{path}}\mathcal{P} - \mathcal{P}i\mathcal{L}_{\mathbf{path}}\mathcal{Q} \neq 0. \tag{A.1}$$

The non-vanishing nature of this commutator implies that paths escaping into the orthogonal subspace $\mathcal{Q}$ undergo trapping with a non-exponential waiting-time distribution, directly defining the non-local core of the semi-Markov renewal kernel, $K(t-\tau)$, capturing long-range memory effects in the degradation kinetics [14].

To track the macroscopic consequences of this transport, we introduce the integrated macroscopic damage potential, Φ. Following the computational principles of continuum damage mechanics [17], for each discrete time layer j (where $t_j = j\Delta t$, the current value of the potential is calculated as a step-by-step time integral over the history of accumulated microstructural defects up to the current layer:

$$\Phi_{j+1} = \Phi_j + \int_{t_j}^{t_{j+1}} f\Big( \big\|[i\mathcal{L}_{\mathbf{path}}, \mathcal{P}]\big\| \cdot n(\tau)\Big) d\tau, \tag{A.2}$$

where n(τ) is the local defect density, and the function f governs the non-linear conversion rate of continuous transport into permanent structural damage. The coupling of the integrated layers ensures that the damage trajectory follows a characteristic sigmoid, or **S-shaped curve**, capturing the three fundamental stages of material degradation: the incubation period (Stage I), the linear steady-state accumulation (Stage II), and the final avalanche breakdown (Stage III) [18, 55].

The transition into Stage III is governed by the state of a **Griffiths absorbing macro-barrier**, denoted by its critical thermodynamic threshold $G_c$. To properly condition the stochastic trajectories on surviving or failing to cross this macro-barrier, we employ the generalized Doob's h-transform [23]. The evolution of the system from the perspective of the critical barrier failure is described by the adjoint Doob's equation, operating in **backward time** (τ ≤ t):

$$-\frac{\partial h(x,\tau)}{\partial \tau} = i\mathcal{L}^{\dagger}_{\mathbf{path}} h(x,\tau) + \int_0^{\tau} K(\tau - s)\mathcal{P}h(x,s)ds + \beta\dot{\Phi}_j \big(G_c - \Phi_j\big)^{-1}, \tag{A.3}$$

where $i\mathcal{L}^{\dagger}_{\mathbf{path}}$ is the adjoint path Liouville operator, $h(x,\tau)$ represents the stochastic weight acting as the value function of conditional survival under the generalized Doob framework, and β is a coupling coefficient.

Equation (3) explicitly links the local structure of macroscopic sinks and the non-local semi-Markov kernel to the global potential of non-linear defect accumulation $\Phi_j$. As $\Phi_j \to G_c$, the singularity in the driving force of the backward evolution accelerates the drop in the survival probability $h(x,\tau)$, forcing the semi-Markov holding times to collapse and driving the system past the Griffiths macro-barrier into macroscopic fracture.

In radiation physics of solid states and kinetic theory of degradation, the radiation dose (particle fluence $\mathcal{F}$ or absorbed dose $D_{\mathbf{rad}}$ is the primary governing parameter. It does not enter the equations abstractly but strictly scales the microscopic sources of defects and transforms the geometry of macroscopic sinks. Within our constructed semi-Markov system of equations, the radiation dose is explicitly integrated into three key points of the mathematical apparatus:

1. **Into the local characteristic of the meso-grain (Defect source equation).**

The radiation dose directly determines the local defect density n(τ) via the displacement generation rate (the dpa parameter — displacements per atom), which is proportional to the particle flux density $\dot{\mathcal{F}}$.
In the integrand expression for the increment of the macroscopic damage potential $\Phi_j$ on time layers, the local defect density is written as: $n(\tau) = \kappa \cdot \mathcal{F}(\tau) = \kappa \int_0^\tau \dot{\mathcal{F}}(s)ds$, where κ is the effective cross-section for the formation of stable Frenkel pairs (vacancies and interstitials) taking into account their intra-cascade annihilation. Thus, the integral over time layers j (Equation (A.2)) directly transforms into an integral over the accumulated radiation dose.

**2. Into the commutator of the Liouville operators and the sink projector (Structure transfer).** The accumulation of the radiation dose changes the geometry of the phase space itself. Under the action of a continuous flux of radiation defects, macroscopic sinks (voids) grow, changing their radii $R_{\mathbf{void}}(\mathcal{F})$.
This implies that the projection operator of macroscopic sinks becomes dose-dependent: $\mathcal{P} = \mathcal{P}(\mathcal{F})$. As a result, the intensity of the commutator generating the non-local memory kernel is explicitly modulated by the dose: $[i\mathcal{L}_{\mathbf{path}}, \mathcal{P}(\mathcal{F})] = \mathcal{Q}(\mathcal{F}) i\mathcal{L}_{\mathbf{path}} \mathcal{P}(\mathcal{F}) - \mathcal{P}(\mathcal{F}) i\mathcal{L}_{\mathbf{path}} \mathcal{Q}(\mathcal{F})$. Physically, this encodes structure transfer: the higher the dose $\mathcal{F}$, the more developed the surface of the sinks is, and the more severely they "penalize" and capture free defect trajectories, transferring them into semi-Markov traps.

**3. Into the singular term of the adjoint Doob's equation (Approach to the barrier).** In Doob's backward time equation (Equation (A.3)), the radiation dose enters through the rate of change of the macro-potential $\dot{\Phi}_j$. Since $\Phi_j$ grows monotonically along an S-shaped trajectory with increasing fluence, the right-hand side of Doob's equation (the penalty functional) is explicitly governed by the dose: $\beta\dot{\Phi}_j\left(G_c - \Phi_j\right)^{-1} \Rightarrow \beta\left(\frac{\partial\Phi}{\partial\mathcal{F}}.\dot{\mathcal{F}}\right)\left(J_{Ic} - \Phi_j(\mathcal{F})\right)^{-1}$.
Here, the dose acts as a dynamic driver:

- At small doses (Stage I, incubation period), $\Phi_j(\mathcal{F}) \ll J_{Ic}$, the penalty is small, and the survival function h is close to unity.
- Upon reaching the critical radiation dose $\mathcal{F}_{\mathbf{crit}}$, the damage potential $\Phi_j$ approaches the invariant Cherepanov–Rice J-integral $J_{I_c}$. The denominator tends to zero, inducing a radiation stochastic explosion—the avalanche-like destruction of the material (Stage III).

**Summary:** The radiation dose $\mathcal{F}$ is a natural evolutionary parameter of the system. In numerical modeling, the physical time t in the Liouville and Feynman–Kac transport equations (and layers j) is completely replaced by the dose variable $\mathcal{F}$, which allows for plotting the kinetics of voids and degradation in axes standardized for radiation materials science: "Associated wear Φ — Radiation dose (dpa — displacements per atom)". Thus, equation (A.1) is replaced by $[i\mathcal{L}_{\mathbf{path}}, \mathcal{P}(\mathcal{F})] = \mathcal{Q}(\mathcal{F}) i\mathcal{L}_{\mathbf{path}} \mathcal{P}(\mathcal{F}) - \mathcal{P}(\mathcal{F}) i\mathcal{L}_{\mathbf{path}} \mathcal{Q}(\mathcal{F}) \neq 0$, and equations (A.2) and (A.3) are replaced by $\Phi_{j+1} = \Phi_j + \int_{\mathcal{F}_j}^{\mathcal{F}_{j+1}} f\left\|[i\mathcal{L}_{\text{path}}, \mathcal{P}(\mathcal{F}')]\right\| J_{\mathbf{meso}}\left(\sigma_{ij}, \rho_d(\mathcal{F}')\right)d\mathcal{F}'$ and

$$-\frac{\partial h(x,\mathcal{F}')}{\partial\mathcal{F}'} = i\mathcal{L}^{\dagger}_{\mathbf{path}} h(x,\mathcal{F}') + \int_0^{\mathcal{F}'} K(\mathcal{F}' - \mathcal{F}'')\mathcal{P}(\mathcal{F}'')h(x,\mathcal{F}'')d\mathcal{F}'' + \beta\left(\frac{\partial\Phi}{\partial\mathcal{F}}\right)_j \left(J_{Ic} - \Phi_j\right)^{-1}.$$